\documentstyle[12pt, psfig, twoside]{article}
\def\references{\section*{References}
\bgroup\parindent=0pt\parskip=.5ex
\def\refpar{\par\hangindent=2em\hangafter=1}}
\def\endreferences{\refpar\egroup}
\def\reference{\relax\refpar}
\title{Theory of magnetized accretion discs driving jets}
\author{Jonathan Ferreira\\ \\ Laboratoire d'Astrophysique de
  l'Observatoire de Grenoble} 
 \date{\today}

\newcommand{\be}{\begin{equation}}
\newcommand{\ee}{\end{equation}}

\begin{document}

\maketitle

\abstract{In this lecture I review the theory of magnetized accretion
  discs driving jets, with a focus on Young Stellar Objects (YSOs). I
  first introduce observational and theoretical arguments in favor of 
  the ``disc wind'' paradigm. There, accretion and ejection are
  interdependent, requiring therefore to revisit the standard picture of
  accretion discs. A simple magnetostatic approach is shown to already
  provide some insights of the basic phenomena. The magnetohydrodynamic
  equations as well as all usual assumptions are then clearly listed.
  The relevant physical mechanisms of steady-state accretion and ejection
  from Keplerian discs are explained in a model independent way. The
  results of self-similar calculations are shown and critically 
  discussed, for both cold and warm jet configurations. I finally provide
  observational predictions and the physical conditions required in YSOs
  discs. The necessity of introducing a magnetospheric interaction between
  the disc and the protostar is briefly discussed.} 

\section{Introduction}

Collimated ejection of matter is widely observed in several astrophysical
objects: inside our own galaxy from all young forming stars and some X-ray
binaries, but also from the core of active galaxies. All these objects
share the following properties: jets are almost cylindrical in shape; the
presence of jets is correlated with an underlying accretion disc
surrounding the central mass; the total jet power is a sizeable fraction of
the accretion power.  

\subsection{Why should jets be magnetized~?}

There are basically two kinds of jet observations: spectra (blue shifted
emission lines) and images (in the same lines for YSOs, or in radio continuum
for compact objects). Most of these images show jets that are
extremely well collimated, with an opening angle of only some degrees.
On the other hand, the derived physical conditions show that jets are
highly supersonic. Indeed, emission lines require a temperature of order 
$T\sim 10^4$ K, hence a sound speed $C_s \sim 10$ km/s while the typical jet
velocity is $v_j\sim 300$ km/s. The opening angle $\theta$ of a ballistic
hydrodynamic flow being simply $\tan \theta = C_s/v_j$, this provides
$\theta\sim 5^o$ for YSOs, nicely compatible with observations. Thus, jets
could well be ballistic, showing up an inertial confinement. But the
fundamental question remains: {\bf how does a physical system produce an
unidirectional supersonic flow~?} Naively, this implies that confinement 
must be closely related to the acceleration process.

This has soon been recognized in the AGN community (where jets were being
observed for quite a long time, see Bridle \& Perley 1984), leading
Blandford \& Rees to propose in 1974 the ``twin-exaust'' model. In this
model, the central source is emitting a spherically symmetric jet, which is
confined and redirected into two bipolar jets by the external pressure
gradient. Indeed, the rotation of the galaxy would probably produce a
disc-like anisotropic distribution of matter around its center. Thus, in
principle, the ejected plasma could be focused towards the axis of rotation
of the galaxy (where there is less matter there) and thereby accelerated
like in a De Laval nozzle. Such an idea was afterwards applied by Cant{\'o}
(1980) and K\"onigl (1982) for YSOs.  However, this model had severe
theoretical drawbacks, related to Kelvin-Helmholtz instabilities quickly
destroying the nozzle. Besides, it does not explain the very origin of the
spherical flow.

But the strongest argument comes from recent observations, like those of
HH30 (see eg. Ray et al. 1996). We know now that jets are already clearly
collimated close to the central star, with no evidence of any relevant
outer pressure. This implies that jets must be self-collimated. In my
opinion, such an observation also rules out the proposition that jets are
collimated by an outer poloidal magnetic field (Spruit et al. 1997).

The only model capable of accelerating plasma along with a self-confinement
relies on the action of a large scale magnetic field carried along by the
jet. In fact, Lovelace and Blandford proposed independently in 1976
that if a large scale magnetic field would thread an accretion disc, then
it could extract energy and accelerate particules (electron positron pairs
in their models). Then, Chan \& Henriksen (1980) showed, using a simplified
configuration, that such a field could indeed maintain a plasma flow
collimated. But it was Blandford \& Payne who, in 1982, produced the first
full calculation of the interplay between a plasma flow (made of electrons
and protons) and the magnetic field, showing both acceleration and
self-collimation. 

\subsection{What are the jets driving sources~?}

To make a long story short, there are three different situations
potentially capable of driving magnetized jets from young forming
stars\footnote{An alternative model is based on some circulation of matter
  during the early infall stages (see Lery et al. 1999 and references
  therein). However, by  construction, such a model is only valid for Class
  0 sources and cannot be used to explain jets from T-Tauri stars. See also
  Contopoulos \& Sauty (2001).}.
\begin{itemize}
\item {\bf the protostar alone}: these purely stellar winds extract their
  energy from the protostar itself (eg. Mestel 1968, Hartmann \& McGregor
  1982, Sauty et al. 1999).
\item {\bf the accretion disc alone}: ``disc winds'' are produced from a large
  radial extension in the disc, thanks to the presence of a large scale
  magnetic field (eg. Blandford \& Payne 1982, Pudritz \& Norman 1983). They
  are fed with both matter and energy provided by the accretion process
  alone.
\item {\bf the interaction zone between the disc and the
    protostar}: these ``X-winds'' are produced in a tiny region around the
  magnetopause between the disc and the protostar (eg. Shu et al. 1994,
  Lovelace et al. 1999, Ferreira et al. 2000). 
\end{itemize}

Purely stellar wind models are less favoured because observed jets carry
far too much momentum. In order to reproduce a YSO jet, a protostar
should be either much more luminous or rotating faster than observations
show (DeCampli 1981, K\"onigl 1986). This leaves us with either disc-winds
or X-winds\footnote{Camenzind and collaborators proposed an ``enhanced''
  version of stellar winds, related to an old idea of Uchida \& Shibata
  (1984). In this picture, a magnetospheric interaction with the accretion
  disc is supposed to strongly modify both jet energetics and magnetic
  configuration, leading to enhanced ejection from the protostar (Camenzind
  1990, Fendt et al. 1995, see Breitmoser \& Camenzind 2000 and
  references therein).}. From the observational point of view, it is very
difficult to discriminate between these two models. In a nice review, Livio
(1997) gathered a number of arguments for the disc 
wind model. The main idea is to look for a model able to explain jets from
quite a lot of different astrophysical contexts (YSOs, AGN, X-ray
binaries). The only ``universal'' ingredient required is an accretion disc
threaded by a large scale magnetic field. Such a paradigm naturally
explains (qualitatively) all accretion-ejection correlations known and is
consistent with every context: young stars (see Cabrit's contribution, this
volume), microquasars (eg. Mirabel \& Rodriguez 1999) and AGN (see
eg. Serjeant et al. 1998, Cao \& Jiang 1999, Jones et al. 2000).

\subsection{Where does this magnetic field come from~?}

Let's face it: we don't know. There are two extreme possibilities. The
first one considers that the field has been advected by the infalling
material, leading to a flux concentration in the inner disc regions. The
second one relies on a local dynamo action in the disc. Most probably, the
answer lies between these two extreme cases.

If the interstellar magnetic field has been indeed advected, the crucial
issue is the amount of field diffusion during the infall. Indeed, if we
take the fiducial values $n \sim 1\mbox{ cm}^{-3}$ and $B\sim 4\ \mu$G 
of dense clouds and use the law $B\propto n^{1/2}$ (Crutcher 1999), we get
a magnetic field at 1 UA ranging from 10 to $10^3$ G! One must then
consider in a self-consistent way the dynamical influence of the magnetic
field, along with matter energy equation and ionization state (see
Lachieze's contribution). This is something extremely difficult and, as far
as I know, no definite result has been yet obtained (see however Ciolek \&
Mouschovias 1995).    

How exactly disc dynamo works is also quite unclear. Dynamo theory remains
intricate, relying on the properties of the turbulence triggered inside the
disc. In our current picture of accretion discs, these are highly turbulent
because of some instability, probably of magnetic origin (see Terquem's
contribution). Such a turbulence is believed to provide means of efficient
transport inside the disc, namely anomalous viscosity, magnetic diffusivity
and heat conductibility. Obviously, small scale (not larger than the disc
vertical scale height), time dependent magnetic fields will then exist
inside accretion discs. But we are interested in the mean flow (hence mean
field) dynamics. So, in practice what does ejection require~? To produce
two opposedly directed jets, there must be a large scale magnetic field in
the disc, which is open and of one of the following topologies:

\indent {\bf Dipolar}: the field threads the disc, with only a vertical
component at the disc midplane, matter being forced to cross the
field lines while accreting (eg. Blandford \& Payne 1982). 

\indent {\bf Quadrupolar}: field lines are nearly parallel to matter,
entering the disc in its plane and leaving it at its surfaces, with only a
non zero azimuthal component at the disc midplane (eg. Lovelace et al. 1987).

Most jet models and numerical simulations assume a dipolar magnetic
configuration, with no justification. In fact, it turns out from the analysis
of disc physics that only the dipolar configuration is suitable for launching
jets from Keplerian accretion discs (see appendix A in Ferreira 
1997). This has been recently confirmed using $\alpha$ dynamo-generated
magnetic fields (Rekowski et al. 2000). This is due to a change of sign
of the $\alpha$ effect across the disc midplane, as observed in numerical
simulations of MHD turbulence in the shearing box approximation
(Brandenburg \& Donner 1997). But a realistic situation requires to treat
both turbulence and the backreaction of the magnetic field in a
self-consistent way. Indeed, jet production is a means of flux leakage,
hence of possible dynamo self-regulation (Yoshizawa \& Yokoi 1993). 
Anyway, this severe issue of dynamo and turbulence lead theorists to simply
assume the existence of a large scale magnetic field. Its' value and
distribution are then either imposed or obtained as conditions for
stationarity in Magnetized Accretion-Ejection Structures (MAES). 

\subsection{The MAES paradigm}

A large scale (mean) magnetic field of bipolar topology is assumed to
thread an accretion disc, allowing ejected plasma to flow along open field
lines. This field extracts both angular momentum and energy from the
underlying disc and transfers them back to the ejected plasma. There have
been numerous studies of magnetized jets (e.g. Blandford \& Payne 1982,
Heyvaerts \& Norman 1989, Pelletier \& Pudritz 1992, Contopoulos \&
Lovelace 1994, Rosso \& Pelletier 1994, Lery et al. 1999 to cite only a
few), but they all suppose that the underlying disc would support the
jets. In all these works, the disc itself was treated as a boundary
condition, people usually assuming a standard viscous accretion disc
(Shakura \& Sunyaev 1973). However, if jets are to carry away the disc
angular momentum, they strongly influence the disc dynamics.

The investigation of MAES, where accretion and ejection are interdependent,
requires a new theory of accretion discs. The relevant questions that must
be addressed by any realistic model of stationary magnetized
accretion-ejection structures are the following: 

\hspace{1cm} {\bf (1)} What are the relevant physical mechanisms inside the
disc~?  

\hspace{1cm} {\bf (2)} What are the physical conditions allowing
accretion and ejection~?

\hspace{1cm} {\bf (3)} Can we relate jet properties to those of the disc~?

To answer these questions, one must take into account the full 2D problem
(not 3D, thanks to axisymmetry) and {\bf not} treat the disc as infinitely
thin as in a standard disc theory. As a consequence, no toy-model has been
able yet to catch the main features of these accretion-ejection structures.
There, the disc accretion rate exhibits a radial variation as matter is
being ejected. The link between accretion and ejection in a MAES can
therefore be measured by the quantity
$ \xi = \frac{d \ln \dot M_a}{d \ln r} $
called the ``ejection index''. This parameter (which can vary in the disc)
measures a local ejection efficiency. In a standard accretion disc,
$\xi=0$ everywhere leading to a constant accretion rate. A complete theory
of MAES must provide the allowed values of $\xi$ as a function of the disc
properties.

\section{Theoretical framework of MAES}

\subsection{A magnetostatic approach}

\subsubsection{The Barlow experiment}

A wheel made of a conducting material is put between the two poles of an
electromagnet. This device produces a magnetic field perpendicular to the
disc, along its axis of rotation. The wheel brushes against mercury
contained in a tank, thereby allowing to close an electric circuit (mercury,
wire connecting the disc axis, the disc itself). In order to check if some
current is flowing, we can put a small lamp. With a crank, we then provide
a rotating motion to the disc and let it evolve. If the magnetic field is
off, nothing happens: the lamp stays mute (no current) and the disc very
gradually slows down. On the contrary, if the magnetic field is present,
the lamp lights on (a current is flowing) and the disc stops very quickly~!
When the disc is finally motionless, the lamp is also off.

The explanation of this phenomenon lies in electromagnetic induction. The
disc is made of a conducting material, meaning that charged 
particules are free to move inside it. When the disc starts to rotate,
these particules drag the magnetic field along with them. The field lines
become then twisted, showing a conversion of mechanical into magnetic
energy. Once the disc has stopped, all the initial mechanical energy has
been converted (here, dissipated into heat along the whole electric
circuit). The magnetic field was just a mediator between different energy
reservoirs. 

\subsubsection{On the importance of currents}

A MAES is an astrophysical Barlow wheel. Indeed, the rotation of a conducting
material in a static magnetic field $B_z$ induces an electric (more
precisely electromotive) field
\begin{equation}
\vec E_m= \vec u \wedge \vec B \simeq \Omega r B_z \vec{e_r}
\end{equation}
where $\Omega$ is the matter angular velocity and $\vec{e_r}$ the radial unit
vector (in cylindrical coordinates). This field produces an electromotive
force $e$, ie a difference of voltage between the inner $r_i$ and outer
$r_e$ disc radii, namely
\begin{equation}
e= \int_{r_i}^{r_e} \vec E_m \cdot \vec{dr} \simeq \Omega_i r_i^2 B_i
\end{equation}
(where the dominant contribution is assumed to arise at the inner
radius). In steady state, two electric circuits can develop (above and
below the disc midplane) with a current $I_{\pm}= e/Z_\pm$ where $Z_\pm$ is
their impedance. There is therefore an available electric power 
\begin{equation}
P= e(I_- + I_+) = \frac{\Omega_i^2 r_i^4 B_i^2}{Z_{eq}}
\end{equation}
where $Z_{eq}$ is the equivalent impedance. Since a current $I= I_- + I_+$
is flowing inside the disc, it becomes affected by the magnetic 
field through the Laplace force $\vec F = \int_{r_i}^{r_e} I \vec{dr} \wedge
\vec B$, which provides a torque acting against the disc rotation. This
result is consistent with Lenz's law, which states that induced currents
work against the cause that gave birth to them.

Now, the existence of such a torque allows matter to fall towards the
central engine, hence to be accreted. Energy conservation implies that the
mechanical power liberated through accretion, namely
\be
P_{acc} = \frac{GM \dot M_a}{2 r_i} \ ,
\ee
where $\dot M_a$ is the accretion rate, must be equal to the electric power
$P$. This imposes an impedance matching  
\begin{equation}
Z_{eq} = \frac{2 r_i^2 B_i^2}{\dot M_a} \sim \mu_o \sqrt{\frac{GM}{r_i}}
\end{equation}
where $\mu_o$ is the vacuum permeability. The rhs estimate used a magnetic
field close to equipartition with the thermal energy inside the disc (see
Sect~5.2.2). The disc accretion rate depends therefore on the global
electric circuit.  

Electric currents  $I_\pm$ flow along the axis, more precisely are
distributed inside the jets. But if we model such current as being carried
by an electric wire, we can estimate the generated magnetic field, namely 
$B_{\phi}(r) = - \mu_o I/2\pi r$. Such a field is negative, consistent with
the shear given by the disc rotation. Because of this field, there is a
Laplace force towards the axis, ie the famous ``hoop stress'' which
maintains the jets collimated.

To summarize:
\begin{enumerate}
\item {\bf Accretion}: a torque due to the magnetic field extracts the disc
  angular momentum. Such a torque is related to the establishment of two
  parallel electric circuits. Their equivalent impedance is directly linked
  to the accretion rate affordable by the structure.
\item {\bf Global energy budget}: a fraction of the accretion power is
  dissipated into Joule heating (disc and jets heating, allowing
  radiation), another is converted into kinetic power carried by the
  jets.
\item {\bf Jet acceleration}: possible due to the conversion of electric
  into kinetic power. Note that if the impedances $Z_+$ and $Z_-$ 
  (which describe all dissipative effects) are different, then two bipolar
  but {\it asymmetric} jets can be produced. 
\item {\bf Jet collimation}: jets do have naturally a self-collimating
  force {\it if they carry a non-vanishing current}. Since the electric
  circuit must be closed, not all magnetic field lines embrace a non-zero
  current. This obviously implies that all field lines anchored in
  the disc cannot be self-collimated (Okamoto 1999).
 \end{enumerate}

Lots of physics can be understood within the framework of
magnetostatics. However, the precise description of a rotating
astrophysical disc, its interrelations with outflowing plasma as well as
the calculation of the asymptotic current distribution inside the jets
(necessary to understand collimation) quite evidently require a fluid
description.  

\subsection{Magnetohydrodynamics}

Magnetohydrodynamics (MHD) describes the evolution of a collisional ionized
gas (a plasma) submitted to the action of an electromagnetic field. Because
the source of the field lies in the motions of charged particules
(currents), the field is intrinsically tied to and dependent of these 
motions. Therefore, because of this high non-linearity, MHD offers an
incredible amount of behaviours.

\subsubsection{From a multicomponent to a single fluid description}

Circumstellar discs and their associated jets are made of dust and gas,
which is composed of different chemical species of neutrals, ions and
electrons. We should therefore make a multicomponent treatment. However,
this is far too complex especially when energetics comes in. On the other
hand, if all components are well coupled (through collisions) a single
fluid treatment becomes appropriate. 

For each specie $\alpha$, we define its numerical density $n_\alpha$, 
mass $m_\alpha$, electric charge $q_\alpha$, velocity $\vec v_\alpha$ and
pressure $P_\alpha$. The equation of motion for each specie writes
\begin{equation}
  \rho_{\alpha} \frac{D\vec{v}_\alpha}{Dt} = -\vec{\nabla}P_{\alpha} - 
  \rho_{\alpha} \vec{\nabla} \Phi_{G} + \sum_\beta \vec{F}_{\beta \alpha} 
  + n_\alpha q_\alpha (\vec{E} + \vec{v}_\alpha \wedge \vec{B})
\end{equation}
where $D./Dt= \partial ./\partial t + \vec{v}_\alpha\cdot \nabla $ is the
Lagrangean derivative, $\rho_\alpha= n_\alpha m_\alpha$, $\Phi_G$ is the
gravitational potential of the central star and $\vec{F}_{\beta \alpha}$ is
the collisional force due to all other species $\beta$. We can define the
``mean'' flow very naturally as  
\begin{eqnarray}
\rho &=& \sum_\alpha n_\alpha m_\alpha \nonumber \\
\rho \vec u &= & \sum_\alpha m_\alpha n_\alpha \vec v_\alpha \nonumber\\
P &= &  \sum_\alpha n_\alpha k_B T \\
\vec J &=& \sum_\alpha n_\alpha q_\alpha \vec v_\alpha \nonumber
\end{eqnarray}
where $\rho$ is the density, $\vec u$ the velocity, $P$ the pressure, $\vec
J$ the current density and $k_B$ the Boltzmann constant. A single fluid
description becomes relevant whenever the plasma is enough collisional. In
such a case, we can safely assume that all species share the same
temperature $T$. Moreover, we assume that any drift between the mean flow
and a specie $\alpha$ is negligible, namely $\|\vec{v}_\alpha - \vec{u}  \|
\ll  \| \vec{u}\| $. Using Newton's second law ($\sum_{\alpha, \beta} \vec
F_{\alpha \beta}= \vec 0$) and local electrical neutrality ($\sum_\alpha
n_\alpha q_\alpha= 0$), we get the usual dynamical equations for one fluid
\begin{eqnarray}
  \frac{\partial \rho}{\partial t} &+& \nabla \cdot \rho \vec u = 0 \\
  \rho  \frac{D \vec{u}}{Dt}   &=&  - \nabla P  -   \rho \nabla \Phi_{G} +
  \vec{J} \wedge \vec{B} 
\end{eqnarray}
by summing over all species $\alpha$. Even if the bulk of the flow is
made of neutrals, they feel the magnetic force through collisions with ions
(mainly) and electrons, $\vec{J} \wedge \vec{B} = (1 + X) (\vec F_{in} + \vec
F_{en})$ where $X = \rho_i/\rho_n$ is the density ratio of ions to neutrals.  

The evolution of the electromagnetic field is described by Maxwell's
equations, namely (in vacuum)
\begin{eqnarray}
\nabla \cdot \vec B &=&0 \\
\nabla \wedge \vec B &=& \mu_o \vec J + \frac{1}{c^2} \frac{\partial \vec E}
{\partial t} \label{eq:ampere}\\
\nabla \cdot \vec E &=& \frac{\rho_*}{\varepsilon_o} \\
\nabla \wedge \vec E &=& - \frac{\partial \vec B} {\partial t}
\label{eq:faraday}
\end{eqnarray}
Faraday's law (Eq.~\ref{eq:faraday}) shows that the strength of the
electric field varies like $E/B \sim L/t \sim U$ where $U$ is a typical
plasma velocity. Now, using Amp\`ere's equation (\ref{eq:ampere}), shows
that the displacement current has an effect of order $(U/c)^2$ only with
respect to real currents ($c$ is the speed of light). Thus, in a
non-relativistic plasma, it can be neglected providing a current density
\begin{equation}
\vec J = \frac{1}{\mu_o} \nabla \wedge \vec B
\end{equation}
directly related to the magnetic field. Under this approximation, electric
charge conservation 
\begin{equation}
\frac{\partial \rho_*}{\partial t} + \nabla \cdot \vec J = 0 
\end{equation}
shows that no charge accumulation is allowed: the first term is also of
order $(U/c)^2$. Therefore $\nabla \cdot \vec J =0$, implying closed
electric circuits.    

Energy conservation of the electromagnetic field writes
\be
\frac{\partial W}{\partial t} + \nabla \cdot \vec S_{MHD} = - \vec J
\cdot \vec E 
\ee
where $W= W_e + W_m= \varepsilon_oE^2/2 + B^2/2\mu_o$ is the
electromagnetic field energy density and  
\be
\vec S_{MHD} = \frac{\vec E \wedge \vec B}{\mu_o}
\ee 
is the Poynting vector, carrying the field energy remaining after
interaction with the plasma (the term $\vec J \cdot \vec E $). Inside the
non-relativistic framework, the energy density contained in the electric
field is negligible with respect to $W_m$ (of order $(U/c)^2$).

\subsubsection{Generalized Ohm's law}

In order to close the above system of equations, we need to know the
electric field $\vec{E}$. Its' expression is obtained from the electrons
momentum equation, consistently with the single fluid
approximation. Namely, we assume that electrons are so light that they
react almost instantaneously to any force, i.e.
\begin{eqnarray}
  \vec 0 = - \nabla P_e  & +& \sum_\beta \vec{F}_{\beta e} 
  - e n_e (\vec{E} + \vec{v}_e \wedge \vec{B})\\
  \vec{E} + \vec u \wedge \vec B & = & (\vec u - \vec v_e) \wedge \vec{B} 
- \frac{\nabla P_e}{e n_e} + \sum_\beta \frac{m_\beta n_\beta}{e n_e} (\vec
  v_\beta - \vec v_e) 
\end{eqnarray}
Now, using (i) the expression of the Lorentz force $\vec J \wedge \vec B$,
(ii) the approximation $\vec J \simeq e n_e (\vec v_i - \vec v_e)$ and
(iii) neglecting the contribution due to the collisions between electrons
and neutrals (with respect to those involving ions), we obtain 
the generalized Ohm's law
\be
\vec{E}\ +\ \vec{u} \wedge \vec{B} = \eta \vec{J}\  +\ \frac{\vec{J}\wedge
  \vec{B}}{e n_e}\ -\ \left ( \frac{\rho_n}{\rho}\right)^2 \frac{(\vec{J}
  \wedge\vec{B}) \wedge \vec{B} }{m_{in} n_i \nu_{in}}\
-\ \frac{\nabla P_e}{en_e} 
\label{eq:omgen}
\ee
where $\eta=(m_{ne}n_n \nu_{ne} + m_{ie}  n_i\nu_{ie})/(en_e)^2$ is the
electrical (normal) resistivity due to collisions, and $m_{\alpha \beta}$
the reduced mass. The first term on the rhs is the Ohm term, the second is
the Hall effect, the third the ambipolar diffusion term and the fourth a
source of electric field due to any gradient of electronic
pressure. Fortunately, all these terms become negligible with respect to
$\vec u \wedge \vec B$ whenever the plasma is well coupled and ionized
($\rho_n \ll \rho$).

\subsubsection{Plasma energy equation}

This is the trickiest equation and can be written in several forms. We
choose to express the internal energy equation, which can be written as
\be
\rho T \frac{D S}{Dt} = \Gamma - \Lambda
\label{eq:Thermo}
\ee
where $S$ is the plasma specific entropy, $\Gamma$ all heating terms and
$\Lambda$ all cooling terms. Note that a transport term (eg. such as
conduction) can cool the plasma somewhere and heat it elsewhere. The MHD
heating rate due to the interaction between the plasma and the
electromagnetic field writes   
\be
\Gamma_{MHD}  = \vec{J} \cdot (\vec{E} + \vec{u}  \wedge \vec{B} )  \simeq
\eta J^2 + \left ( \frac{\rho_n}{\rho}\right )^2 \frac{ \left| \vec{J}
    \wedge \vec{B} \right |^2 }{m_{in} n_i \nu_{in}} 
\ee
The first term is the Joule effect while the second is the heating due to
ambipolar diffusion. Note that, although dynamically negligible in discs,
such an effect might be responsible for jets heating (Safier 1993, Garcia et
al. submitted). 

\subsection{Modelling a MAES}

\subsubsection{Assumptions}

Our goal is to describe an accretion disc threaded by a large scale
magnetic field of bipolar topology. In order to tackle this problem, we
will make several symplifing assumptions: 

{\bf (i) Single-fluid MHD}: matter is ionized enough and all species well
coupled. As a consequence, we use the simple form of Ohm's law \be
\vec{E} + \vec{u} \wedge \vec{B} = \eta \vec{J} \ee Taking the curl of
this equation and using Faraday's law, we get the induction equation
providing the time evolution of the magnetic field
\begin{equation}
  \frac{\partial \vec B}{\partial t} = \nabla \wedge (\vec u \wedge \vec
  B) - \nabla \wedge (\nu \nabla \wedge \vec B)
  \label{eq:ind1}
\end{equation}
where $\nu= \eta/\mu_o$ is the magnetic diffusivity. The first term
describes the effect of advection of the field by the flow while the
second describes the effect of diffusion, matter being able to cross
field lines thanks to diffusivity.

{\bf (ii) Axisymmetry}: using cylindrical coordinates ($r$, $\phi$,
$z$) no quantity depends on $\phi$, the jet axis being the vertical
axis. As a consequence $E_\phi =0$ and all quantities can be decomposed
into poloidal (the ($r$, $z$) plane) and toroidal components, eg. $\vec u =
\vec{u_p} + \Omega r \vec{e_\phi}$ and $\vec B= \vec{B_p} + B_{\phi}
\vec{e_\phi}$. A bipolar magnetic configuration can then be described with 
\be
\vec{B_p} = \frac{1}{r} \nabla a \wedge \vec{e_\phi}  \ ,
\label{eq:flux}
\ee 
\noindent where $a (r,z)$ is an even function of $z$ and with an odd
toroidal field $B_\phi(r,-z)= -B_\phi(r,r)$. The flux function $a$ is 
related to the toroidal component of the potential vector ($a=rA_{\phi}$)
and $a(r,z)= constant$ describes a surface of constant vertical magnetic
flux $\Phi$,
\be
\Phi= \int_S \vec B\cdot \vec{dS} = 2\pi a(r,z) \ .
\ee
The magnetic field distribution in the disc as well as the total amount of
flux are unknown and must therefore be prescribed.

{\bf (iii) Steady-state}: all astrophysical jets display proper
motions and/or emission nodules, showing that they are either prone to some
instabilities or that ejection is an intermittent process. However, the
time scales involved in all objects are always much larger than the
dynamical time scale of the underlying accretion disc. Therefore, a steady
state approach is appropriate\footnote{Note that this conclusion holds even in
  microquasars. In GRS 1915+105 the duration of an ejection event is around
  $10^3$ sec only, but the disc dynamical time scale is around 1 msec, ie.
  $10^{-6}$ times smaller. See however Tagger \& Pellat (1999) for an
  alternative view on this topic.}.   

{\bf (iv) Transport coefficients}: if we use a normal (collisional)
value for $\nu$, we find that the ratio of advection to diffusion in
(\ref{eq:ind1}), measured by the magnetic Reynolds number 
\be
{\cal R}_m \sim \frac{L U}{\nu} \ ,
\ee
is always much larger than unity inside both the disc and the jets. This is
the limit of ideal MHD where plasma and magnetic fields are ``frozen
in''. Jets are therefore described with ideal MHD ($\eta=0$). Within this
framework, the stronger (initially the magnetic field) carries the weaker
(ejected disc plasma) along with it. But inside the disc, gravitation is
the dominant energy source and the plasma drags and winds up the field
lines. Such a situation cannot be maintained for very long. Instabilities
of different kind will certainly be triggered leading to some kind of
saturated, turbulent, disc state (eg. tearing mode instabilities, or
magneto-rotational instability, Balbus \& Hawley 1991).
 
In turbulent media, all transport effects are enhanced, leading to
anomalous transport coefficients. These coefficients are the magnetic
diffusivity $\nu_m$ and resistivity $\eta_m$, but also the viscosity
$\nu_v$ (associated with the transport of momentum) and thermal
conductivity $\kappa_T$ (associated with the heat flux). Providing the
expressions of these anomalous coefficients requires a theory of MHD
turbulence inside accretion discs. Having no theory, we will use simple
prescriptions, like the alpha prescription used by Shakura \& Sunyaev
(1973). In particular, because of the dominant Keplerian motion in discs,
we allow for a possible anisotropy of the magnetic diffusivity. Namely,
we use two different turbulent diffusivities: $\nu_m$ (related to the
diffusion in the ($r$, $z$) plane) and $\nu'_m$ (related to the diffusion
in the $\phi$ direction).

Since stationary discs must be turbulent, all fields (eg. velocity,
magnetic field) must be understood now in some time average sense.   

{\bf (v) Non-relativistic MHD framework}: in YSOs, matter remains always
non-relativistic but there is something more about it. Indeed, magnetic
field lines are anchored in an rotating accretion disc. Hence, if a field
line of angular velocity $\Omega_*$ opens a lot, there is a cylindrical
distance such that its linear velocity reaches the speed of light,
defining a ``light cylinder'' $R_L= c/\Omega_*$. Now, if one imposes
ideal MHD regime along the jet, ie. $\vec E= - \vec u \wedge \vec B =
-\Omega_* r \vec{e_{\phi}} \wedge \vec{B_p}$, we see that the
displacement current is no longer negligible at the light cylinder:
propagation effects become relevant and one must take into account a local
non-zero electric charge $\rho_*$ (provided by the Goldreich-Julian charge
$\nabla \cdot \vec E = \rho_*/\varepsilon_o$). As a consequence, the plasma
feels an additional electric force $\rho_* \vec E$, even if the bulk
velocity of the flow is non relativistic (eg. Breitmoser \& Camenzind
2000)! 

But remember that this extra force and its corresponding ``light
cylinder'' arose because of the {\it assumption} of ideal MHD. In fact,
taking into account a local charge density probably imposes to also treat
the non-ideal contributions (see Eq.~\ref{eq:omgen}). Nobody provided yet a
self-consistent calculation. We just assume here that any charge
accumulation would be quickly canceled (no dynamical relevance of the
light cylinder).

\subsubsection{Set of MHD equations}

We use the following set of MHD equations:

\noindent {\bf Mass conservation}
\be 
\nabla \cdot \rho \vec u  = 0
\label{1}
\ee
\noindent {\bf Momentum conservation}
\be 
\rho \vec u \cdot \nabla \vec u = -\nabla P - \rho \nabla \Phi_G 
+ \vec J \wedge \vec B  +  \nabla \cdot {\mathsf T}
\label{2}
\ee
where $\mu_o \vec J= \nabla \wedge \vec B$ is  the current density and
${\mathsf T}$ the turbulent stress tensorm which is related to the turbulent
viscosity $\nu_v$ (Shakura \& Sunyaev 1973). 

\noindent {\bf Ohm's law} and {\bf toroidal field
  induction}\footnote{Obtained from Eq.~(\ref{eq:ind1}) and after some
  algebra (remember that $E_\phi=0$).}  
\begin{eqnarray}
  \eta_m \vec J_{\phi}  &=& \vec{u_p} \wedge \vec{B_p}\label{3}\\
  \nabla \cdot (\frac{\nu'_m}{r^2}\nabla rB_{\phi}) & = &
  \nabla \cdot \frac{1}{r}(B_{\phi} \vec{u_p} - \vec{B_p}\Omega r)
  \label{4} 
\end{eqnarray}
where $\eta_m=\mu_o \nu_m$ and $\eta'_m= \mu_o\nu'_m$ are the anomalous 
resistivities. 

\noindent {\bf Perfect gas law}
\be
P= \rho \frac{k_B}{\mu m_p} T
\label{5}
\ee
where $m_p$ is the proton mass and $\mu$ a generalized ``mean molecular
weight''.  

\noindent {\bf Energy equation}\\
As seen previously, the exact energy equation (\ref{eq:Thermo}) involves
various physical mechanisms. Its explicit form is 
\be
\nabla \cdot (U \vec{u_p}\ +\ \vec S_{rad}\ +\ \vec q_{turb}) =  -P \nabla
\cdot \vec{u_p}\ +\ \eta_m J^2_{\phi}\ +\ \eta'_mJ^2_p\ +\ \eta_v
\left| r \nabla \Omega \right|^2 
\label{eq:ener}
\ee
where $\eta_v= \rho \nu_v$, $U$ is the internal specific energy, $\vec
S_{rad} = -\frac{c}{\kappa \rho} \nabla P_{rad}$ is the radiative energy
flux ($\kappa$ is the Rosseland mean opacity of the 
plasma, $P_{rad}$ the radiation pressure) and $\vec q_{turb}$ the unknown
turbulent energy flux. This flux of energy arises from turbulent motions
and provides both a local cooling $\Lambda_{turb}$ and heating
$\Gamma_{turb}$. Indeed, using a kinetic description and allowing for
fluctuations in the plasma velocity and magnetic field, it is possible to
show that all energetic effects associated with these fluctuations cannot
be reduced to only anomalous Joule and viscous heating terms. Therefore, a
consistent treatment of turbulence imposes to take into account
$\vec q_{turb}$ (see early paper of Shakura et al. 1978). But how to do
it~? Moreover, the radiative flux depends on the local opacity $\kappa$ of
the plasma, which varies both radially and vertically inside the
disc. Besides, the expression of the radiative pressure $P_{rad}$ is known
only in optically thick media ($P_{rad}= a T^4$), while the disc can be
already optically thin at the disc midplane.

Thus, it seems that a realistic treatment of the energy equation is still
out of range. As a first step to minimize its impact on the whole
structure, we will use a polytropic equation   
\be 
P= K \rho^{\Gamma} \ ,
\label{6}
\ee 
\noindent where the polytropic index $\Gamma$ can be set to vary between 1
(isothermal case) and $\gamma=\frac{5}{3}$ (adiabatic case) for a
monoatomic gas. Here $K$ can be allowed to vary radially but remains
constant along each field line. In section~5.3, we will turn our attention
to thermal effects and use therefore a more appropriate form of the energy
equation. 

Using the above set of MHD equations to describe astrophysical discs and
jets would normally require consistency checks, at least {\it a posteriori}.
If one fluid MHD seems justified inside the inner regions of accretion discs, 
it may well not be anymore the case along the jet (huge fall in
density). In particular non-ideal effects are most probably starting to
play a role in Ohm's law (ambipolar diffusion or even Hall terms), allowing
matter to slowly drift across field lines. This may have important
dynamical effects dowstream. To clearly settle this question, the full
thermodynamics of the plasma including its ionization state should be
self-consistently solved along the jet. 

\subsection{Critical points in stationary flows}

In the real world, everything is time dependent. Imagine matter expelled
from the disc without the required energy: it will fall down, thereby
modifying the conditions of ejection. A steady state is eventually reached
after a time related to the nature of the waves travelling upstream and
providing the disc information on what's going on further up.

The adjustment of a MAES corresponds to the phenomenon of impedance
matching. As we saw, this matching relates the accretion rate to the 
dissipative effects in the electric circuit. In practice, this means that
the resolution of stationary flows requires to take into account, in some
way, all time-dependent feedback mechanisms. This is done by requiring
that, once a steady-state is achieved, no information (ie. no waves) can
propagate upstream, from infinity (in the z-direction) to the accretion
disc. There is only one way to do it. Matter must flow faster than any
wave, leaving the disc causally disconnected from its surroundings. 

\subsubsection{The Parker wind}

Let us first look at the simple model of a spherically symmetric,
isothermal, hydrodynamic flow. Such a model was first proposed by Parker
(1958) to explain the solar wind. In spherical coordinates, mass
conservation and momentum conservation write
\begin{eqnarray}
\frac{du}{u} + \frac{d\rho}{\rho} + 2\frac{dr}{r} &=& 0\\
u \frac{du}{dr} + \frac{1}{\rho}\frac{dP}{dr} + \frac{GM}{r^2} &=& 0
\end{eqnarray}
where gas pressure is $P= \rho C_s^2$, the sound speed $C_s$ being a
constant. In such a simple system is hidden a singularity. Indeed, after
differentiation one gets
\begin{eqnarray}
(C_s^2 - u^2)\frac{d\ln \rho}{dr} &=& 2 \frac{u^2}{r} -\frac{GM}{r^2} \\
(C_s^2 - u^2)\frac{d\ln u}{dr} &=& -2 \frac{C_s^2}{r} +\frac{GM}{r^2}
\end{eqnarray}
showing that the system is singular at $r=r_s$, where $u=C_s$. In order to
obtain a stationary solution, one must then impose the regularity condition
which is
\be
r_s = \frac{GM}{2C_s^2}
\ee
In practice, fixing the distance of the sonic point imposes the value of
one free parameter (eg. the initial velocity or density). Only the
trans-sonic solution is stationary.

\subsubsection{Critical points of a MAES}

In a magnetized medium, there are several waves able to transport
information, related to the different restoring forces. In a MAES, the Lorentz
force couples with the plasma pressure gradient leading to three different
MHD waves: 
\begin{itemize}
\item the Alfv\'en wave (A), causing only magnetic disturbances along the
  unperturbed magnetic field $B_o$ and of phase velocity
  \[V_A = \frac{B_o}{\sqrt{\mu_o \rho}} \] 
\item two magnetosonic waves, the slow (SM) and the fast (FM), involving
  both magnetic disturbances and plasma compression (rarefaction), of
  phase velocity 
  \[V^2_{SM, FM} = \frac{1}{2} \left ( V_A^2 + C_s^2 \mp \sqrt{ (V_A^2 +
      C_S^2)^2 - 4 V_A^2 C_s^2 \cos^2\theta} \right ) \]
  where $\theta$ is the angle between the unperturbed field $B_o$ and the
  direction of propagation of the wave (the disturbance).
\end{itemize}
In ideal MHD regime (in the jets), these three waves can freely
propagate. This provides three singularities along {\it each} magnetic
surface, whenever the plasma velocity equals one of these phase
speeds. Thus, three regularity conditions must be specified per magnetic
surface.   

Inside the turbulent accretion disc it is another story. There, the high
level of turbulence maintains large magnetic diffusivities and
viscosity. As a result, MHD waves are strongly damped and the number of
singularities really present is not so clear. Within the alpha prescription
of accretion discs (Shakura \& Sunyaev 1973), viscosity appears only in the
azimuthal equation of motion (angular momentum conservation). The presence
of viscosity there ``damps'' the acoustic waves and there is no singularity
related to this motion (although rotation is supersonic). Conversely, there
is no viscosity in the poloidal (radial and vertical) equations of motion:
a singularity can therefore appear there, related to pure acoustic waves. As
a consequence, in principle, it may be necessary to impose a regularity
condition with respect to the accretion flow itself, if it becomes
supersonic.

\section{Magnetized jets}

\subsection{Commonly used equations}

As said previously, jets are in ideal MHD regime, namely
$\nu_v=\nu_m=\nu'_m=0$. In this regime, mass and flux conservations 
combined with Ohm's law (\ref{3}) provide  
\be  
\vec{u}_p = \frac{\eta (a)}{\mu_o\rho} \vec B_p
\label{MHDI1}
\ee 
\noindent where $\eta(a)=\sqrt{\mu_o\rho_A}$ is a constant along a magnetic
surface\footnote{Any quantity $Q$ verifying $\vec B_p \cdot \nabla Q=0$
  is a constant along a poloidal magnetic field line, hence a MHD invariant on
  the corresponding magnetic surface.} and $\rho_A$ is the density at the
Alfv\'en point, where the poloidal velocity $u_p$ reaches the poloidal
Alfv\'en velocity $V_{Ap}$. The induction equation (\ref{4}) becomes 
\be 
\Omega_*(a) = \Omega -\eta\frac{B_{\phi}}{\mu_o\rho r} \ ,
\label{MHDI2}
\ee 
\noindent where $\Omega_*(a)$ is the rotation rate of a magnetic
surface (imposed by the disc, thus very close to the Keplerian value). Note
that despite the frozen in situation, jet plasma flows along a magnetic
surface with a total velocity $\vec  u = (\eta/\mu_o\rho) \vec B\ 
+\ \Omega_*r \vec{e_\phi}$ which is {\bf not} parallel to the total magnetic
field $\vec B$. This is possible because field lines rotate faster than the
ejected plasma. If the disc is rotating counter-clockwise, the field lines
will be trailing spirals ($\Omega_* > \Omega$, ie. $B_\phi <0$ with
$B_z>0$), while ejected plasma will rotate in the same direction as in the
disc. Magnetic field lines and plasma stremlines are thus two helices of
different twist. Jet angular momentum conservation simply writes
\be 
\Omega_*r^2_A = \Omega r^2 - \frac{rB_{\phi}}{\eta}
\label{MHDI3}
\ee 
\noindent where $r_A$ is the Alfv\'en radius. Above the disc, the turbulent
torque vanishes and only remains a magnetic accelerating torque. The first
term on rhs is the specific angular momentum carried by the ejected plasma
whereas the last tern can be understood as the angular momentum stored in
the magnetic field. The total specific angular momentum $L(a)=
\Omega_*r^2_A$ is an MHD invariant. The Alfv\'en radius $r_A$ can be
interpretated as a magnetic lever arm, braking down the disc. The larger the
ratio $r_A/r_o$, the larger the magnetic torque acting on the disc at the
radius $r_o$. 

Hereafter, we focus only on adiabatic jets (for which thermal effects can
still be non negligible). Usually, instead of using the other two
components of the momentum conservation equation, one uses the Bernoulli
equation (obtained by projecting Eq.~(\ref{2}) along the poloidal
direction, ie. $\vec B_p$) and the transverse field or Grad-Shafranov
equation (obtained, after quite a lot of algebra, by projecting
Eq.~(\ref{2}) in the direction perpendicular to a magnetic surface,
ie. $\nabla a$). For a jet of adiabatic index $\gamma$, Bernoulli equation
writes 
\begin{equation}  
  E(a)= \frac{u^2}{2} + H + \Phi_G - \Omega_* r\frac{B_{\phi}}
  {\sqrt{\mu_o\rho_A}} 
  \label{eq:bern}
\end{equation} 
where $u$ is the total plasma velocity and $H= (\gamma/\gamma-1)P/\rho$ is
the gas enthalpy. This equation describes the acceleration of matter along
a poloidal magnetic surface, namely the conversion of magnetic energy and
enthalpy into ordered kinetic energy. Grad-Shafranov equation of an adiabatic
jet is
\begin{eqnarray}
  \nabla \cdot (m^2 -1) \frac{\nabla a}{\mu_o r^2}  &= & \rho \left \{
    \frac{d E}{d a} - \Omega \frac{d \Omega_*r_A^2}{d a} \right. \ +\  
  \left . (\Omega r^2 - \Omega_* r^2_A)\frac{d \Omega_*}{d a}\ -\
    \frac{C_s^2}{\gamma(\gamma-1)} \frac{d \ln K}{d a} \right \}
  \nonumber\\ 
  & & \ +\  \frac{B^2_{\phi} + m^2B^2_p}{\mu_o} \frac{d \ln \eta}{d a}  
  \label{eq:GS}
\end{eqnarray}
where $m^2 \equiv u^2_p/V^2_{Ap}$ is the Alfv\'enic Mach number and $C_s^2= 
\gamma k_B T/\mu m_p$ is the jet sound speed. This awful equation
provides the transverse equilibrium (ie. the degree of collimation) of a
magnetic surface. A simpler-to-use and {\bf equivalent} version of this
equation is  
\be
(1-m^2) \frac{B^2_p}{\mu_o {\cal R}}\  -\  \nabla_{\perp}\left(P +
\frac{B^2}{2\mu_o}\right)\ -\ \rho \nabla_{\perp} \Phi_G\ +\ (\rho\Omega^2r -
\frac{B^2_{\phi}} {\mu_o r} )\nabla_{\perp}r = 0
\label{eq:fperp}
\ee
where $\nabla_{\perp}\equiv \nabla a \cdot \nabla /|\nabla a | $
provides the gradient of a quantity perpendicular to a magnetic surface
($\nabla_{\perp} Q <0$ for a quantity $Q$ decreasing with increasing magnetic 
flux) and ${\cal R}$, defined by
\be
\frac{1}{{\cal R}} \equiv \frac{\nabla a}{|\nabla a|} \cdot 
\frac{(\vec{B}_p\cdot\nabla)\vec{B}_p}{B^2_p} \ ,
\ee
is the local curvature radius of a particular magnetic surface. When ${\cal
R} >0$, the surface is bent outwardly while for ${\cal R} <0$, it bends
inwardly. The first term in Eq.(\ref{eq:fperp}) describes the reaction to
the other forces of both magnetic tension due to the magnetic surface (with
the sign of the curvature radius) and inertia of matter flowing along it
(hence with opposite sign). The other forces are the total pressure
gradient, gravity (which acts to close the surfaces and deccelerate the
flow, but whose effect is already negligible at the Alfv\'en surface), and
the centrifugal outward effect competing with the inwards hoop-stress due
to the toroidal field.

To summarize, an astrophysical jet is a bunch of axisymmetric magnetic
surfaces $a(r,z)= Const.$, nested one around the other at different anchoring
radii $r_o$. The magnetic flux distribution $a(r_o)$ is unknown and is
therefore prescribed, whereas the shape $r(z)$ of the magnetic surface is
self-consistently calculated. Each magnetic surface is characterized by 5 MHD
invariants: 

\hspace{1cm}- $\eta(a)$, ratio of ejected mass flux to magnetic flux;

\hspace{1cm}- $\Omega_*(a)$, the rotation rate of the magnetic surface;

\hspace{1cm}- $L(a)= \Omega_* r_A^2$, the total specific angular momentum
transported;

\hspace{1cm}- $E(a)$, the total specific energy carried away;

\hspace{1cm}- $K(a)$, related to the specific entropy $S(a)$.

A magnetized jet is then described by 8 unknown variables: density $\rho$,
velocity $\vec u$, magnetic field $\vec B$ (flux function $a$ and toroidal
field $B_\phi$), pressure $P$ and temperature $T$. There are 8 
equations allowing us to solve the complete problem: (\ref{eq:flux}),
(\ref{5}), (\ref{6}), (\ref{MHDI1}), (\ref{MHDI2}), (\ref{MHDI3}),
(\ref{eq:bern}) and (\ref{eq:GS}) or its more physically meaningful version
(\ref{eq:fperp}). Since the ejected plasma must become super-FM, 3
regularity conditions have to be imposed, leaving the problem with 5 free
boundary conditions (at each radius $r_o$). Studies of disc-driven jets
usually assume that matter rotates at the Keplerian speed, namely
$\Omega_o= \Omega_K(r_o)= \sqrt{GM/r_o^3}$ and that jets are ``cold''
(negligible enthalpy, $K(a)=0$) or choose an arbitrary distribution
$K(a)$. In both cases, it leaves the problem with only 3 {\bf free and
  independent} boundary conditions that must be specified at each
radius\footnote{In numerical MHD computations, one usually prescribes the
  density $\rho^+(r_o)$, vertical velocity $u^+_z(r_o)$ and magnetic field
  $B^+_z(r_o)$ distributions (eg. Ouyed \& Pudritz 1997, Krasnopolsky et
  al. 1999). In those self-similar jet studies where only the Alfv\'en
  point has been crossed, 4 (if $K(a)= 0$, Blandford \& Payne 1982) of 5
  (if $K(a)\neq 0$, Contopoulous \& Lovelace 1994) variables remain free and 
  independent ($\rho^+$, $u_z^+$, $B^+_z$, $B^+_\phi$ and $P^+$). In both
  cases, the superscript ``+'' stands here for the {\bf belief} that this
  boundary condition corresponds indeed to the disc surface.}.

Magnetized jets are such complicated objects that only gross properties are
known. For example, we know that a non-vanishing asymptotic current will
produce a self-confinement of {\it some} field lines (Heyvaerts \& Norman
1989). But the exact proportion of collimated field lines depends on
``details'' (transverse gradients of inner properties, outer
pressure). The distance at which jets become collimated, the jet radius and
opening angle, the velocity and density transverse distributions still
remain to be found in full generality. This is the reason why there are so
many different works on jet dynamics and why each authors use their own
``relevant'' parameters. Following Blandford \& Payne (1982), we introduce
the following jet parameters 
\begin{eqnarray}
\lambda & = &\frac{\Omega_*r^2_A}{\Omega_or^2_o} \simeq 
\frac{r^2_A}{r^2_o} \simeq 1  - \frac{B_\phi^+}{\eta \Omega_o r_o}\nonumber\\ 
\kappa & = &\eta \frac{\Omega_or_o}{B_o} \simeq \frac{\mu_o\Omega_o
  r_o}{B_o^2}\rho^+ u_z^+ 
\end{eqnarray}
The index ``o'' refers to quantities evaluated at the disc midplane, ``+''
at the disc surface and ``A'' at the Alfv\'en point. The first parameter,
$\lambda$, is a measure of the magnetic lever arm that brakes down the disc
while $\kappa$ measures the ejected mass flux. Another parameter is usually
introduced, related somehow to the jet asymptotic behaviour. We use  
\be
\omega_A = \frac{\Omega_*r_A}{V_{Ap,A}} \simeq \kappa \lambda^{1/2}
\frac{B_o}{B_{p,A}}
\ee
which measures the ratio of the rotational velocity to the poloidal jet
velocity at the Alfv\'en point. Such a parameter characterises the magnetic
rotator: cold jets require  $\omega_A >1$ to become trans-Alfv\'enic
(Pelletier \& Pudritz 1992, Ferreira 1997, Lery et al. 1999). Note also
that its value depends on pure geometrical effects, namely the way the
magnetic surface opens. As a consequence, spherical expansion of field
lines is probably a special case, leading to particular relations between
jet asymptotic behaviour and its source.

\subsection{Some aspects of cold jets physics}

\subsubsection{Energetic requirements}

It is quite reasonable to assume that the magnetic energy density in the
disc is much smaller than the gravitational energy density. As a
consequence, the rotation of the disc drags the magnetic field lines which
become then twisted. This conversion of mechanical to magnetic energy in
the disc gives rise to an outward poloidal MHD Poynting flux 
\begin{equation}
\vec{S}_{MHD,p} = \frac{\vec{E}\times \vec{B}_\phi}{\mu_o}= - \Omega_* r
B_{\phi} \frac{\vec{B}_p}{\mu_o} 
\end{equation}
which feeds the jets and appears as the magnetic term in the Bernoulli
integral (\ref{eq:bern}). Another source of energy for the jet could be the
enthalpy $H$ (built up inside the disc and advected by the ejected flow) or
another local source of heating ${\cal Q}$ (eg. coronal heating). We are
mainly interested here in ``cold'' jets, where those two terms are
negligible (see Sect.~5.3). Bernoulli equation can be rewritten as
\be
E(a)= \frac{u_p^2}{2}\ +\ \Phi_{eff}\ +\ H
\ee
where the effective potential is $\Phi_{eff}= \Phi_G - \frac{1-g^2}{2}
\Omega_*^2 r^2$ with $\Omega= \Omega_*(1-g)$ and $\Omega_* \simeq
\Omega_o$. The function $g$ is much smaller than unity at the disc surface
then increases along the jet (if the jet widens a lot, $g \rightarrow
1$). Starting from a point located at the disc surface $(r_o, z=0)$, matter
follows along a magnetic surface and must move to another point $(r_o +
\delta r_o, z)$. This can be done only if a positive poloidal velocity is
indeed developed. Making a Taylor expansion of $\Phi_{eff}$, one gets
$\frac{u_p^2}{2} \simeq H_o - H + \frac{\Omega_o^2}{2}(3 \delta r_o^2 -
z^2) >0$, which translates into the condition
\be
\tan \theta = \frac{z}{\delta r_o} < \sqrt{3} \left ( 1 + \frac{2}{3}
  \frac{H_o - H}{\Omega_o^2 \delta r_o^2} \right )^{1/2} \ .
\ee 
Thus, cold jets (negligible enthalpy) require field lines bent by more than
$30^o$ with respect to the vertical axis at the disc surface (Blandford \&
Payne 1982). The presence of a significant enthalpy ($H_o$ large) is
obviously required if this condition is not satisfied. Bernoulli equation
(\ref{eq:bern}) gives a total energy feeding a {\bf cold} jet
\be
E(a)= \frac{\Omega_o^2 r_o^2}{2} - \Omega_o^2 r_o^2 + \Omega_*(\Omega_*
r_A^2 - \Omega_o r_o^2) = \frac{\Omega_o^2 r_o^2}{2} (2\lambda - 3)
\ee
which is directly controlled by the magnetic lever arm $\lambda$. Cold,
super-FM jets require therefore $\lambda >3/2$. If all available energy is
converted into kinetic energy, ejected plasma reaches an asymptotic
poloidal velocity $u_\infty(a) = \sqrt{2 E(a)}$.

\subsubsection{Relevant forces and current distributions}

So, rotation of open field lines produces a shear ($B^+_\phi <0$) that
results in an outward flux of energy. Once matter is loaded onto 
these field lines, it will be flung out whenever there are forces
overcoming the gravitational attraction. Since matter flows along magnetic
surfaces, one must look at the projection of all forces along these
surfaces. We obtain for the Lorentz force, 
\begin{eqnarray}
F_{\phi} & = & \frac{B_p}{2\pi r} \nabla_{\parallel} I \nonumber \\
F_{\parallel} & = & - \frac{B_{\phi}}{2\pi r}\nabla_{\parallel} I
\end{eqnarray}
where $\nabla_{\parallel}= \vec{\nabla} \cdot \vec{B}_p/B_p$ and $I= 2\pi r
B_{\phi}/\mu_o$ is the total poloidal current flowing inside the magnetic
surface. Hence, jets are magnetically-driven whenever $\nabla_{\parallel} I
> 0$ is fulfilled, namely when current is leaking through this
surface. This quite obscur condition describes the fact that magnetic
energy is being converted into kinetic energy: the field lines accelerate
matter {\bf both} azimuthaly ($F_{\phi} >0$) and along the magnetic surface
($F_{\parallel}>0$). The difference with the Barlow wheel lies in the
possibility to convert magnetic energy into jet (bulk) kinetic energy.

The jet transverse equilibrium depends on the subtle interplay between several
forces (see Eq.~(\ref{eq:fperp})). The transverse projection of the
magnetic force, namely 
\be
F_{\perp} = B_pJ_{\phi} - \frac{B_{\phi}}{2\pi r} \nabla_{\perp} I
\ee
where $\nabla_{\perp} \equiv (\nabla a \cdot\nabla)/|\nabla a|$, shows that
it depends on the transverse current distribution. Thus, the degree of jet
collimation (as well as plasma acceleration) depends on the overall
electric current circuit. Any bias introduced on this circuit can produce
an artificial force and modify diagnostics on jet collimation.

Eventually matter becomes no longer magnetically accelerated and
$\nabla_{\parallel} I= 0$ is satisfied. This implies two possible
asymptotic current distributions. If jets are force-free (ie. $\vec J_p$
and $\vec B_p$ parallel), then there is a non-vanishing asymptotic current
providing a self-collimating pinch (and one must worry about how the
electric circuit is closed). Or, jets become asymptotically current-free
($I_\infty= 0$) and another cause must then be responsible for their
collimation (either inertial or external pressure confinement). This last
alternative has something appealing for magnetic fields would then have a
major influence only at the jet basis, becoming dynamically negligible
asymptotically (cf Sect.~1.1).  

\subsection{Numerical simulations: what can be learned of them~?}

There have been a lot of numerical studies of MHD jet propagation and their
associated instabilities. Here, I focus only on those addressing the
problem of jet formation from accretion discs. Although some attempts have
been made to model accretion discs driving jets, difficulties are such that
nothing realistic has been obtained yet (Shibata \& Uchida 1985, Stone \&
Norman 1994, Kudoh et al. 1998). Ejection is indeed observed, but no one
can tell whether these events are just transients or if they indeed
represent some realistic situation. 

Another philosophy is to treat the disc as a boundary condition and,
starting from an (almost) arbitrary initial condition, wait until the system
converges towards a stationary flow (Ustyugova et al. 1999, 2000, Ouyed \& 
Pudritz 1997, 1999, Krasnopolsky et al. 1999). Now, it is of no wonder that
jets are indeed obtained with these simulations. Matter, forced to flow
along open magnetic field lines, is continuously injected (at a rate
$\rho^+ u^+_z$) at the bottom of the computational box. As a response, the
field lines twist (ie. $B_{\phi}$ increases) until there is enough magnetic
energy to propell it. If the code is robust enough, a steady-state
situation is eventually reached, actually reproducing most of the results
obtained with self-similar calculations (Krasnopolsky et al. 1999). Ouyed
\& Pudritz (1997) found a parameter region where unsteady solutions are
produced, with a ``knot generator'' whose location seems to remain
fixed. Their results may be related to the characteristic recollimation
configuration featured by self-similar solutions (Blandford \& Payne 1982,
Contopoulos \& Lovelace 1994, Ferreira 1997).

Nevertheless, since the time step for computation depends on the faster waves
(Courant condition) whose phase speed varies like $\rho^{-1/2}$, it appears
that no current MHD code can cope with tiny mass fluxes. Thus, code
convergence itself introduces a bias in the mass flux of numerical
jets. These are always very ``heavy'', with a magnetic lever arm $r_A/r_o$
not even reaching 4 (Ouyed \& Pudritz 1999, Krasnopolsky et al. 1999,
Ustyugova et al. 1999) while cold self-similar studies obtained much larger
values (with a minimum value around 10). This is not a limitation imposed
by physics but of computers and will certainly be overcome in the future. 

On the other hand, the question of jet collimation is clearly one that can
be addressed by simulations. However, the boundary conditions imposed at
the box, as well as the shape of the computational domain, can introduce
artificial forces leading to unsteady jets or spurious collimation (see the
nice paper of Ustyugova et al. 1999).

Anyway, the following crucial question remains to be addressed: {\bf how
  (and how much) matter is loaded into the field lines~?} Or another way to
put it: how is matter steadily deflected from its radial motion (accretion)
to a vertical one (ejection)~? To answer this question one must treat the
accretion disc in a self-consistent way.

\section{Magnetized accretion discs driving jets}

\subsection{Physical processes in quasi-Keplerian discs}

\subsubsection{Turbulent, Keplerian discs}

As said previously, we focus our study on Keplerian accretion discs. Such a
restriction imposes negligible radial plasma pressure gradient and
magnetic tension. We define the local vertical scale height as $P_o=
\rho_o \Omega_K^2 h^2$ where $P_o$ is the total plasma pressure. Looking at
the disc radial equilibrium, a Keplerian rotation rate is indeed obtained
whenever the disc aspect ratio
\be
\varepsilon = \frac{h(r)}{r}
\ee
is smaller than unity. Hereafter, we use $\varepsilon <1$ as a free
parameter and we will check {\it a posteriori} the thin disc approximation
(Sect.~5.4.1).

Steady accretion requires a turbulent magnetic diffusivity for matter must
cross field lines while accreting and rotating. Since rotation is much
faster than accretion, there must be a higher dissipation of toroidal field
than poloidal one. {\it A priori}, this implies a possible anisotropy of
the magnetic diffusivities associated with these two directions, poloidal
$\nu_m$ and toroidal $\nu_m'$. Besides, such a turbulence might also
provide a radial transport of angular momentum, hence an anomalous
viscosity $\nu_v$. To summary, at least three anomalous transport
coefficients are necessary to describe a stationary MAES. We will use the
following dimensionless parameters defined at the disc midplane:
\begin{eqnarray}
  \alpha_m = \frac{\nu_m}{v_A h} & & \mbox{ level of turbulence}\nonumber \\
  \chi_m = \frac{\nu_m}{\nu_m'} & & \mbox{ degree of anisotropy}\\
  {\cal P}_m = \frac{\nu_v}{\nu_m} & & \mbox{ magnetic Prandtl number}
  \nonumber  
\end{eqnarray}
where $v_A= B_o/\sqrt{\mu_o \rho_o}$ is the Alfv\'en speed. Our
conventional view of 3D turbulence would translate into $\alpha_m < 1$,
$\chi_m \sim 1$ and ${\cal P}_m \sim 1$. But as stated before, the amount
of current dissipation may be much higher in the toroidal direction,
leading to $\chi_m \ll 1$. Moreover, it is not obvious that $\alpha_m$ must
necessarily be much smaller than unity. Indeed, stationarity requires that
the time scale for a magnetic perturbation to propagate in the vertical
direction, $h/v_A$, is longer than the dissipation time scale,
$h^2/\nu_m$. This roughly translates into $\alpha_m > 1$. Thus, we must be
cautious and will freely scan the parameter space defined by these turbulence
parameters.

\subsubsection{How is accretion achieved~?}

The disc being turbulent, accretion of matter bends the poloidal field
lines whose steady configuration is provided by Ohm's law (\ref{3}). At the
disc midplane, this equation provides
\be
{\cal R}_m \equiv \frac{r u_o}{\nu_m} = \left . \frac{\mu_o r J_\phi}{B_z}
\right |_{z=0} \sim \frac{l^2}{r^2}
\ee
where ${\cal R}_m$ is the magnetic Reynolds number related to the
radial motion $u_o$ and $l(r)$ is the characteristic scale height of the
magnetic flux variation. Once the disc turbulence properties are given, we
just need to know the accretion velocity $u_o$. This is provided by the
angular momentum equation, namely
\begin{equation}
\rho \frac{\vec{u}_p}{r} \cdot \nabla \Omega r^2 = F_{\phi} +
(\nabla \cdot {\mathsf T} )\cdot \vec e_\phi
\end{equation}
where $F_{\phi}$ is the magnetic torque due to the large
scale magnetic field (``jet'' torque) and $(\nabla \cdot {\mathsf T} )\cdot
\vec e_\phi$ is the ``viscous''-like torque of turbulent origin (possibly
due to the presence of a small scale magnetic field). Such an equation can
be put into the following conservative form
\be
\nabla \cdot \left [ \rho \Omega r^2 \vec u_p\ -\ \frac{r
    B_\phi}{\mu_o}\vec B_p\ -\ r \vec T_v \right ] = 0
\ee
where $r(\nabla \cdot {\mathsf T} )\cdot \vec e_\phi = \nabla \cdot r \vec
T_v$. Although modelling the jet torque is quite straightforward, it is not
the case of the turbulent torque $\vec T_v$ and we use a prescription
analogous to Shakura \& Sunyaev (1973). Defining 
\begin{equation}
  \Lambda = \left | \frac{\mbox{jet torque}}{\mbox{turbulent torque}} \right
  |_{z=0} 
\end{equation}
the disc angular momentum conservation becomes at the disc midplane
\be
1 + \Lambda \simeq {\cal R}_m \left (\frac{\nu_m}{\nu_v}\right) =
\frac{{\cal R}_m }{{\cal P}_m }
\ee
Taking the conventional value ${\cal P}_m \sim 1$, one sees that a
``standard'' accretion disc, which is dominated by the viscous torque
($\Lambda \ll 1$) requires straigth poloidal field lines (${\cal R}_m \sim
1$, Heyvaerts et al. 1996). On the contrary, cold jets carrying
away all disc angular momentum ($\Lambda \gg 1$) are produced with bent
magnetic surfaces  so that ${\cal R}_m \sim \Lambda \sim \varepsilon^{-1}$
(Ferreira \& Pelletier 1995). 

\begin{figure}[h]
\centerline{\hbox{\psfig{file=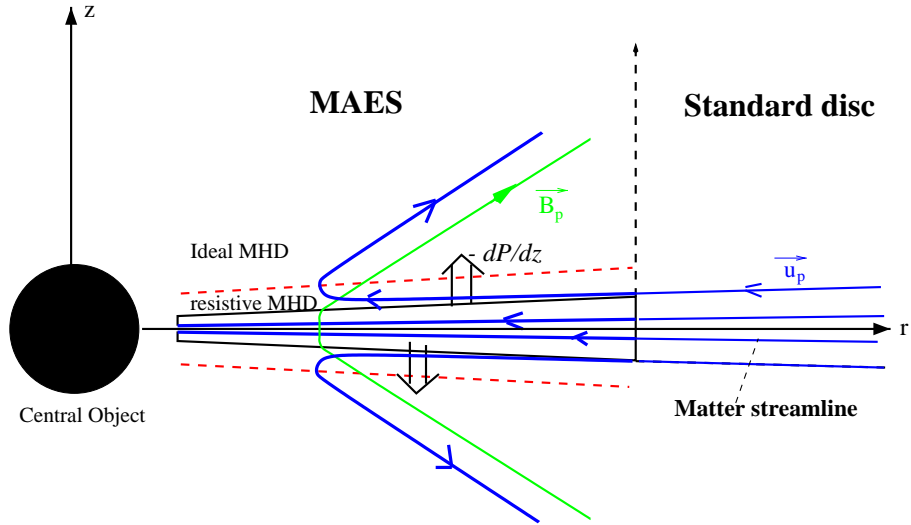,width=12cm}}}
\caption[]{Sketch of a Magnetized Accretion Ejection Structure (Courtesy of
  Fabien Casse).}
\end{figure}

\subsubsection{How is matter defleted from accretion~?}

The poloidal components of the momentum conservation equation write
\begin{eqnarray}
  \left( \vec{u}_p \cdot \vec{\nabla} \right) u_r &=& (\Omega^2 - \Omega_K^2)r
  + \frac{F_r}{\rho} - \frac{1}{\rho} \frac{\partial P}{\partial r} \\
  \left( \vec{u}_p \cdot \vec{\nabla} \right) u_z &\simeq & - \frac{1}{\rho}
  \frac{\partial P}{\partial z} - \Omega_K^2 z
  - \frac{1}{\rho} \frac{\partial}{\partial z} \frac{B^2_r + B^2_{\phi}}{2
  \mu_o} 
\label{eq:pz}
\end{eqnarray}
At the disc midplane, a total (magnetic + viscous) negative torque provides
an angular velocity slightly smaller than the Keplerian one
$\Omega_K=\sqrt{GM/r_o^3}$, thereby producing an accretion motion
($u_r<0$). But at the disc surface, one gets an outwardly directed flow
($u^+_r>0$) because both the magnetic tension $F_r$ (whose effect is
enhanced by the fall in density) and the centrifugal force overcome gravitation
($\Omega^+ > \Omega_K$). Note that jets are magnetically-driven, the
centrifugal force resulting directly from the positive magnetic torque
($F^+_{\phi} >0$). This can be understood with a simple geometrical argument:
matter has been loaded onto field lines that are anchored at inner radii
and are thus rotating faster.

But again, we assumed that matter is being loaded from the underlying
layers with $u^+_z >0$. The physical mechanism is hidden in
Eq.~(\ref{eq:pz}): the only force that can {\it always} counteract both
gravity and magnetic compression is the plasma pressure gradient
(Ferreira \& Pelletier 1995). Inside the disc, a quasi-MHS equilibrium is
achieved, with matter slowly falling down ($u_z <0$) while
accreting. However, plasma coming from an outer disc region eventually
reaches the upper layers at inner radii. There, the plasma pressure
gradient slightly wins and gently lifts matter up ($u^+_z >0$), at an
altitude which depends on the local disc energetics.

\subsubsection{How is steady ejection obtained~?}

Is ejection unavoidable once all above\footnote{Namely, rotation, open
  field lines and some amount of diffusion allowing loading of matter.}
ingredients are met~? The answer is ``yes'', but {\bf steady} ejection
requires another condition.  

While accretion is characterized by a negative azimuthal component of
the Lorentz force $F_{\phi}$, magnetic acceleration occurring in jets
requires a positive $F_{\phi}$. Since $F_{\phi}= J_zB_r -
J_rB_z$, the transition between these two situations depends mainly on the
vertical profile of the radial current $J_r \equiv - \mu_o^{-1}
\partial B_{\phi}/\partial z$, that is, on the rate of change of the
magnetic shear with altitude. In order to switch from accretion to
ejection, $J_r$ must vertically decrease on a disc scale height. This
crucial issue is controlled by Eq.~(\ref{4}), that provides  
\be
\eta'_m J_r \simeq \eta'_oJ_o\ +\ r\int_0^zdz\vec{B_p}\cdot \nabla\Omega\
-\ B_{\phi} u_z 
\ee
The first term on the rhs describes the current due to the electromotive
force, the second is the effect of the disc differential rotation and the
third is the advection effect, only relevant at the disc surface
layers. Thus, the vertical profile of $J_r$ is mainly controlled by the
ratio of the differential rotation effect over the induced current
(Ferreira \& Pelletier 1995). No jet would be produced without
differential rotation for it is the only cause of the vertical decrease of
$J_r$. However, the counter current due to the differential rotation cannot
be much bigger than the induced current in the disc, otherwise $J_r$ would
become strongly negative and lead to an unphysically positive toroidal
field at the disc surface. Thus, steady state ejection is achieved
only when these two effects are comparable, which translates into
\begin{equation}
\Lambda \sim \Lambda_c \equiv \frac{3 \chi_m}{\alpha_m^2 {\cal P}_m
  \varepsilon} 
\label{eq:Ga}
\end{equation}
For $\Lambda > \Lambda_c$, matter is spun down at the disc surface, while
for $\Lambda < \Lambda_c$ there is not enough energy to propell the large
amount of mass trying to escape from the disc. Thus, equation~(\ref{eq:Ga})
is a necessary condition for stationarity.

\subsubsection{From resistive discs to ideal MHD jets}

As matter is expelled off the disc (by the plasma pressure gradient) with
an angle $\theta_{u_p} \equiv \arctan(u_r/u_z)$, magnetic stresses make it
gradually flow along a magnetic surface (with $\vec{u_p} \parallel
\vec{B_p}$). Indeed, Ohm's law~(\ref{3}) can be written 
\be
\eta_m J_{\phi} = u_r B_z \left( {{\tan \theta_{B_p}}\over{\tan 
\theta_{u_p}}} -1 \right )
\ee
where $\theta_{B_p} \equiv \arctan(B_r/B_z)$. As long as
$\theta_{u_p} < \theta_{B_p}$, the toroidal current remains
positive. This maintains a negative vertical Lorentz force (that
decreases $u_z$) and a positive radial Lorentz force (that, along with
the centrifugal term, increases $u_r$), thus increasing
$\theta_{u_p}$. If $\theta_{u_p} > \theta_{B_p}$, the toroidal current
becomes negative, lowering both components of the poloidal Lorentz
force and, hence, decreasing $\theta_{u_p}$. Therefore, in addition to the
vertical decrease of the magnetic diffusivity (for its origin lies in
a turbulence triggered inside the disc), there is a natural mechanism
that allows a smooth transition between resistive to ideal MHD regimes.

\subsection{Disc-jets interrelations}

\subsubsection{Dimensionless parameters}

The accretion disc is defined by 11 variables, ie. the same 8 as in the
ideal MHD jets plus the 3 transport coefficients. All these quantities can
be calculated from their values at the disc equatorial plane. At a
particular radius $r_o$ they write 
\[
\begin{array}{lcl}
  P_o  =  \rho_o \Omega_K^2 h^2 & &\\ 
  T_o = \frac{G M m_p}{k_B r_o} \varepsilon^2 & & \\
  u_{r,o} = - u_o= - m_s \varepsilon \Omega_K r_o  &\mbox{      where    }&
  m_s = 2 q \mu \frac{1 + \Lambda}{\Lambda}= \alpha_v \varepsilon (1 +
  \Lambda)\\ 
  \left . \frac{d u_z}{dz} \right |_{z=0} = \frac{u_o}{r}(\xi -1) & & \\
  \Omega_o = \delta\, \Omega_K &\mbox{      where    }& \delta = \left( 1 -
    \varepsilon^2 \left [ \frac{m_s^2}{2} + 2(2- \beta) + \mu {\cal R}_m
    \right ] \right)^{1/2} \\
  \rho_o = \frac{\dot M_{ao}}{4\pi \Omega_K r_o^3 m_s \varepsilon^2}& & \\  
  B_o = \left(\frac{\mu}{m_s}\right)^{1/2} \left ( \frac{\mu_o \dot M_{ao}
      \Omega_K}{4 \pi r_o} \right)^{1/2}
  &\mbox{      where    }& \mu= \frac{B_o^2}{\mu_o P_o}\\
  \left . \frac{d B_\phi}{dz} \right |_{z=0} = -\mu_o J_{r,o} = - q
  \frac{B_o}{h}  &\mbox{      where    }& q = 
  \frac{\alpha_m {\cal P}_m \Lambda \varepsilon}{2\mu^{1/2}} \delta \\
  J_{\phi,o} = {\cal R}_m \frac{B_o}{\mu_o r} &\mbox{      where    }&
  {\cal R}_m = {\cal P}_m (1 + \Lambda)
\end{array}
\]
where $\nu_v = \alpha_v \Omega_K h^2$, $\Omega_K= \sqrt{GM/r_o^3}$, $\dot
M_{ao}= \dot M_a(r_o)$, $\beta= d\ln a/d\ln r_o$ provides the magnetic flux
distribution and $\Lambda$ is constrained by Eq.~(\ref{eq:Ga}). Thus, there
are 3 more parameters in addition to the previous 4 ones ($\varepsilon$,
$\alpha_m$, ${\cal P}_m$, $\chi_m$): the magnetic flux distribution
$\beta$, the magnetic field strength $\mu$ and the ejection index
$\xi$. The 3 regularity conditions arising at the 3 critical points met by
the ejected plasma along each magnetic surface provide the value of 3 disc
parameters (precisely, their values at the anchoring radius $r_o$). Inside
our cold approximation, we therefore expect to fix the values of $\beta$,
$\mu$ and $\xi$ as functions\footnote{An important remark. The field
  strength cannot be too large ($\mu >>1$), otherwise there will be no
  vertical equilibrium possible. On the other hand, if it is too small ($\mu
  <<1$), then the disc is prone to the magneto-rotational instability (Balbus
  \& Hawley 1991). Therefore, we expect $\mu \sim 1$ in steady-state
  MAES. However, the allowed values of $\xi$ strongly depend on the (subtle)
  vertical equilibrium and its interplay with the induction equation.}  of
the free parameters $\varepsilon$, $\alpha_m$, ${\cal P}_m$ and
$\chi_m$. As a consequence, jet properties (ie. invariants as well as the
asymptotic behaviour) arise as by-products of these parameters.

Using the set of ideal MHD equations, mass conservation gives a mass flux
leaving the disc surface $\rho^+ u_z^+ \simeq \xi \varepsilon \rho_o u_o$
related by $\xi$ to the accretion mass flux. Then, angular momentum
conservation provides the following exact relations 
\begin{eqnarray}
  \lambda &=& 1 + \frac{\Lambda}{2\xi(1 + \Lambda)} \left |
    \frac{B_\phi^+}{qB_o} \right | \nonumber \\ 
  \kappa &=&  \frac{q}{\lambda - 1} \left | \frac{B_\phi^+}{qB_o} \right |
  \label{qlamb}
\end{eqnarray}
Both magnetic lever arm $\lambda$ and mass load $\kappa$ are therefore
determined by the ejection index $\xi$ (which is directly related to the
exact value of the toroidal field $B_{\phi}^+$ at the disc
surface\footnote{The estimate $B_{\phi}^+ \simeq - qB_o$ is roughly acurate
  (by less than a factor 2) but the jet asymptotic structure highly depends
  on its precise value (Ferreira 1997).}). One way to understand this point
is to compute the ratio $\sigma$ of the MHD Poynting flux to the kinetic
energy flux. At the disc surfaces this ratio writes
\be
\sigma_+ = \left | \frac{ - \Omega_* r B_\phi \vec B_p }{\mu_o \frac{\rho
      u^2}{2} \vec u_p} \right |_+ = 2(\lambda -1) = \frac{1}{\xi}
 \frac{\Lambda}{1+\Lambda} \left | \frac{B_\phi^+}{qB_o} \right |
\ee
The ejection index appears to be also a measure of the power feeding the
jets. Unless $\xi$ is of order unity, the magnetic field completely
dominates matter at the disc surface (Eq.~(\ref{eq:Ga}) forbids $1 >>
\Lambda \sim \xi$).   

\subsubsection{Constraints on the ejection index of cold MAES}

Can we provide any general constraint on the allowed range of $\xi$~? This
is the trickiest question about MAES and here follows some analytical
arguments in the simple case of cold jets from isothermal discs (Ferreira
1997).  

The maximum value $\xi_{max}$ is constrained by the jet capability to
accelerate ejected matter, hence momentum conservation. Let's consider two
extreme cases. Imagine there is such a tiny fraction of mass ejected that
it takes almost no energy to accelerate it until the Alfv\'en point. But as
some acceleration has been provided anyway, we can write $\sigma_A <
\sigma_+$. On the other extreme, a huge amount of matter is expelled off
the disc, which has hardly enough energy. In this most extreme case, no
more energy is left after the Alfv\'en point and $m^2_{max}= 1$. Gathering
these two conditions provides the constraint
\be
\frac{1 + 2 \xi}{1 - 4\xi} < \omega_A^2 < \frac{1}{2\xi}
\ee
which shows that cold jets (1) require $\omega_A >1$ (fast rotators) and (2)
display a maximum ejection efficiency of $\xi_{max} = \frac{\sqrt{13} -
  3}{4}\simeq 0.15$. Higher ejection indices are inconsistent with
steady-state, trans-Alfv\'enic, cold jets.

The minimum value $\xi_{min}$ of the ejection index is constrained by the
disc vertical equilibrium. The diminishing of the ejection efficiency $\xi$
is obtained by increasing the magnetic compression, especially through the
radial component (via $\lambda$, increasing as $\xi$ decreases). Now, to
maintain the vertical balance while bending the field lines, but without
increasing the plasma pressure, one must decrease the magnetic field
amplitude (parameter $\mu$). But then, there is a non-linear feedback on
the toroidal field induction. Indeed, as $\mu$ decreases, the effect due to
the differential rotation decreases also, leading to an increase of the
toroidal field at the disc surface. This causes an increase of the toroidal
magnetic pressure, hence an even greater magnetic squeezing of the
disc. Below $\xi_{min}$, no vertical equilibrium is possible. Providing a
quantitative analytical expression of how much matter can actually be
ejected, ie. the value of $\xi_{min}$, is out of range. Indeed, in 
the resistive upper disc layers, {\bf all dynamical terms are comparable} in
Eq.~(\ref{eq:pz}). A careful treatment of the disc vertical balance is
therefore badly needed. Thus, finding the correct parameter space of a MAES
forbids the use of crude approximations, like $\rho u_z=constant$
(Wardle \& K\"onigl 1993), using strict hydrostatic balance (Li 1995)
or any other prescription mimicking the induction equation (Li 1996).

\subsubsection{Jet asymptotic structure}

Can we relate the jet asymptotic structure to the disc parameters without
solving the full set of MHD equations~? Naively, one would say that the
larger $\lambda$ the larger jet asymptotic radius, {\it if some cylindrical
  collimation is achieved}. Next section, we will see that this last issue
is far from being obvious. Anyway, we can still safely say that if some
current is still available after the Alfv\'en point, then magnetic
acceleration will probably occur, along with an opening of the magnetic
surfaces. In fact, the larger $B_\phi^+$, the larger $B_{\phi,A}$. This, in
turn, ensures that jets will provide a big acceleration and open up a
lot. For a cold jet, the ratio $I_A/I_+$ of the remaining current $I_A$ to the
current provided at the disc surface $I_+$ is 
\be
\frac{I_A}{I_+}= \frac{r_A B_{\phi,A}}{r_o B_{\phi}^+} \simeq g_A\ \ \ \
\mbox{   where   }\ \ \ \  g_A^2 = 1 - \frac{3}{\lambda} 
- \frac{1}{\omega^2_A} + \frac{2}{\lambda^{3/2}(1+z^2_A/r^2_A)^{1/2}} 
\ee
The expression of $g_A$ is the Bernoulli equation evaluated at the Alfv\'en
point and 
\be 
\omega_A \simeq \kappa \lambda^{3/2} \frac{\sin
  (\phi_A-\theta_A)}{\sin\phi_A} 
\label{eq:Wa}
\ee
where $\phi_A$ is the local angle between the Alfv\'en surface (defined by
$z=z_A(r_A)$) and the vertical axis, and $\theta_A$ the opening angle
estimated at the Alfv\'en point. Both angles are determined by the
resolution of the Grad-Shafranov equation, which takes into account radial
boundary conditions. Note that this expression is only valid for a conical
Alfv\'en surface, a geometry which arises naturally when jet parameters
vary slowly across the jet. Since cold jets require fast magnetic rotators
($\omega_A >1$), a necessary condition for trans-Alfv\'enic jets is $\kappa
\lambda^{3/2} >1$, which translates into $\lambda > (\Lambda \alpha_v
\varepsilon)^{-2}$.  Thus, jets launched from discs with a dominant viscous
torque $\Lambda <1$ require huge magnetic lever arms, namely $\lambda
\gg \varepsilon^{-2}$. This is most probably forbidden by the disc vertical
equilibrium that would not survive such a strong magnetic pinching. So,
cold disc-driven jets are presumably carrying away a significant fraction
of the disc angular momentum ($\Lambda > 1$).

\subsubsection{Summary}

From the preceeding general analysis, we can {\it a priori} expect two
extreme {\bf cold} configurations from quasi-Keplerian discs:  
\begin{description}
\item[Type I,] where large toroidal currents $J_\phi$ at the disc midplane
  correspond to large magnetic Reynolds numbers ${\cal R}_m \sim
  \varepsilon^{-1} >> 1$ and a dominant magnetic torque $\Lambda \sim
  \varepsilon^{-1}$. This configuration would be achieved for isotropic
  turbulence, ${\cal P}_m \sim 1$ and $\chi_m \sim 1$. 
\item[Type II,] where the dominant source of toroidal currents is at the
  disc surfaces, corresponding to straight field lines inside the disc (${\cal
    R}_m \sim 1$) that become bent only at the surface. These surface
  currents come from the electromotive force ($J_\phi^+ \simeq - (u_r
  B_z/\eta_m)^+$), due to the presence of a large viscous torque ($\Lambda
  \sim 1$) allowing a non zero accretion velocity at the disc surface. Such
  a configuration would be achieved for an anisotropic turbulence, ${\cal
  P}_m > 1$ and $\chi_m \sim \varepsilon$.   
\end{description}
At this stage, I hope the reader has achieved an understanding of the
relevant physical mechanisms inside a keplerian accretion disc driving
jets (question 1, Sect.~1.4). The disc physical conditions (question 2) 
are described by the MAES parameter space and thus, require the
treatment of the complete set of MHD equations. As a ``by-product'', we
will hopefully have the answer of the last question (jet properties).

\section{Self-similar solutions of MAES}

\subsection{Self-similar Ansatz and numerical procedure}

Solving the full set of MHD equations requires heavy 2D or 3D numerical
simulations. However, looking for special solutions will allow us to
transform the set of partial differential equations (PDE) into two sets of
ordinary differential equations (ODE) with singularities. Gravity is
expected to be the leading energy source and force in accretion
discs. Thus, if MAES are settled on a wide range of disc radii, magnetic
energy density probably follows the radial scaling imposed by the
gravitational energy density. The gravitational potential writes in
cylindrical coordinates 
\be
\Phi_G(r,z) = - \frac{GM}{r} \left (  1 + \frac{z^2}{r^2} \right )^{-1/2}
\ee 
Since the disc is a system subjected to the dominant action of gravity, any
physical quantity $A(r,z)$ will follow the same scaling, namely $A(r,z) =
G_A(r) f_A(\frac{z}{r})$. Since gravity is a power law of the disc radius,
we use the following self-similar Ansatz 
\be
A(r,z)= A_e\left(\frac{r}{r_e}\right)^{\alpha_A}f_A(x)
\ee
where $x = z/h(r)= z/\varepsilon r$ is our self-similar variable and $r_e$
is the MAES outer radius. Because all quantities have power law
dependencies, the resolution of the ``radial'' set of equations is trivial
and provides algebraic relations between all exponents. The most general
set of radial exponents allowing to take into account {\bf all} terms in
the dynamical equations (ie. no energy equation) is:
\begin{eqnarray*}
\beta = \frac{3}{4}\ +\ \frac{\xi}{2} &\ \ \ \ & \alpha_\rho = \xi\ -\
\frac{3}{2}\\ 
\alpha_{B_r} = \alpha_{B_\phi} = \alpha_{B_z} = \beta\ -\ 2 &\  \  \  \ &
\alpha_P = \alpha_\rho\ -\ 1 \\ 
\alpha_{u_r}= \alpha_{u_\phi}= \alpha_{u_z}=\ - \frac{1}{2} &\ \ \ \ &
\alpha_{\nu_m} = \alpha_{\nu'_m} = \alpha_{\nu_v} = \frac{1}{2} 
\end{eqnarray*}
As an illustration, the solutions obtained by Blandford \& Payne used
$\beta=3/4$, ie $\xi =0$. Note also that the disc scale height must verify
$h(r)= \varepsilon r$. Such a behaviour stems only from dynamical
considerations, ie. the vertical equilibrium between gravity and magnetic
compressions and plasma pressure gradient. However, the energy equation
provides another constraint that is usually incompatible with such a
scaling (see eg. Ferreira \& Pelletier 1993). We will come back to this
issue later on. 

All quantities $f_A(x)$ are obtained by solving a system of ODE which can
be put into the form
\[ \left (\begin{array}{ccc} \ldots & & \\ &\bf M &\\ & & \ldots
  \end{array} \right) 
\cdot \left (\begin{array}{c} \frac{df_1}{dx} \\ \vdots \\ \frac{df_n}{dx}
  \end{array} \right)
= \left (\begin{array}{c} \ldots \\ \bf P \\ \ldots \end{array} \right) \]
where ${\bf M}$ is a 8x8 matrix in resistive MHD regime, 6x6 in ideal MHD
(see Ferreira \& Pelletier 1995). A solution is therefore available
whenever the matrix ${\bf M}$ is inversible, namely its determinant is
non-zero. Starting in resistive MHD regime, $det\, {\bf M}=0$ whenever
\be
V^2(V^2 - C_s^2) = 0 \ ,
\ee
where $C_s$ is the sound speed and $V \equiv \vec u \cdot \vec n$ is the
critical velocity. The vector 
\be
\vec n = \frac{ \vec e_z - x \varepsilon \vec e_r} {(1 + x^2
  \varepsilon^2)}
\ee
provides the direction of propagation of waves that are consistent with our
axisymmetric, self-similar description. Therefore, close to the disc, the
critical velocity is $V\simeq u_z$, whereas far from the disc it becomes
$V\simeq u_r$ (no critical point in the azimuthal direction). Thus, inside
the resistive disc, the anonalous magnetic resistivities produce such a
dissipation (presence of high order derivatives) that the magnetic 
force does not act as a restoring force and the only relevant waves are
sonic. Note also that the equatorial plane where $V=0$ is also a critical
point (of nodal type since all the solutions must pass through it). This
introduces a small difficulty, since one must then begin the integration
slightly above it. In the ideal MHD region,  $det \, {\bf M}=0$ whenever
\be
(V^2- V^2_{SM})(V^2-V^2_{FM})(V^2 - V_{An}^2)^2=0
\ee
namely, where the flow velocity $V$ successively reaches the three phase
speeds $V_{SM}$, $V_{An}$ and $V_{FM}$, corresponding respectively to the
slow magnetosonic wave, the Alfv\'en wave and the fast magnetosonic 
wave. The phase speeds of the two magnetosonic modes have the usual
expression, namely $V^2_{SM,FM}= \frac{1}{2} \left ( C_s^2 + V_{At}^2 \mp
  \sqrt{(C_s^2 + V_{At}^2)^2 - 4C_s^2V_{An}^2}  \right ) $ where $V_{At}$
is the total Alfv\'en speed and $V_{An}= \vec V_{Ap} \cdot \vec n$. Note
that the condition $V=V_{An}$ is equivalent to $u_p=V_{Ap}$, which shows
that this is the usual Alfv\'enic critical point encountered in 
jet theories. It is also noteworthy to remark that the multiplicity of this
root implies that at the Alfv\'enic point, both first and second order
derivatives of the physical quantities are imposed by the regularity
condition.   

How do we proceed~? Starting slightly above the disc midplane where all
quantities are known, we propagate the resistive set of equations using a
Runge-Kutta (better a Stoer-Burlisch) solver. We do this for fixed values
of the four free parameters $(\varepsilon, \alpha_m, \chi_m, {\cal P}_m)$
and some guesses for $\mu$ and $\xi$. As $x$ increases, the flow reaches an
ideal MHD regime and we shift to the corresponding set of equations. Care
must be taken in order to not introduce jumps in the solution while doing
this. If, for the chosen ejection efficiency $\xi$, the value $\mu$ of the
magnetic field is too large, the overwhelming magnetic squeezing leads to a
decrease of $u_z$. If, on the contrary, $\mu$ is too small, the plasma
pressure gradient becomes far too efficient and leads to infinite vertical
acceleration. Using these two criteria and fine-tunning $\mu$, we can approach
the SM point so close that a simple linear extrapolation on all quantities
allow us to safely cross the singularity. This leapfrog must not introduce
any discontinuity (to some tolerance) on all jet invariants. By doing so,
we obtain a trans-SM solution. This is done for a chosen value of $\xi$
that may not be the critical $\xi_c$ allowing a trans-A solution. If $\xi <
\xi_c$, the magnetic tension wins over the outwardly directed tension
produced by the ejected, rotating plasma. As a result, the magnetic surface
closes which leads to a decceleration and the Alfv\'en point is not
reached. On the other hand, $\xi > \xi_c$ produces an over-opening of the
magnetic surface leading to the unphysical situation where $B_\phi$ goes to
zero. Again, once close to the Alfv\'en point (typically, $m=1$ by 1\%), we
do a leapfrog. Thus, fine-tunning $\xi$ and, for each guess, finding the
critical value of $\mu$, allow us to obtain trans-SM and super-Alfv\'enic
solutions. No trans-FM solution {\it connected to the disc} has been found yet.

\subsection{``Cold'' configurations}

A cold configuration is defined by a negligible enthalpy at the disc
surface (at each magnetic surface). Since discs are quasi-keplerian, cold
jets are obtained with an isothermal (Wardle \& K\"onigl 1993, Ferreira \&
Pelletier 1995, Li 1995, 1996, Ferreira 1997) or adiabatic (Casse \&
Ferreira 2000a) {\it vertical} profile $f_T(x)$ of the temperature.

\subsubsection{General behaviour}

The general behaviour is exactly that expected in the disc. Both magnetic
and (if non negligible) turbulent viscous torques extract energy and
angular momentum from disc plasma. The accretion velocity depends on the
amount of magnetic diffusivity. As plasma is being accreted from the outer
disc regions, it is slightly converging towards the disc midplane (because of
both tidal and magnetic compressions). Matter located (locally) at the disc
surface feels a strong outwardly directed magnetic tension and a positive
azimuthal torque, both arising from current consommation ($\nabla_\parallel
I >0$): accretion is stopped and reversed. More or less simultaneously, a
positive vertical velocity is provided by the plasma pressure gradient.

\begin{figure}
  \begin{tabular}{cc}
    \psfig{file=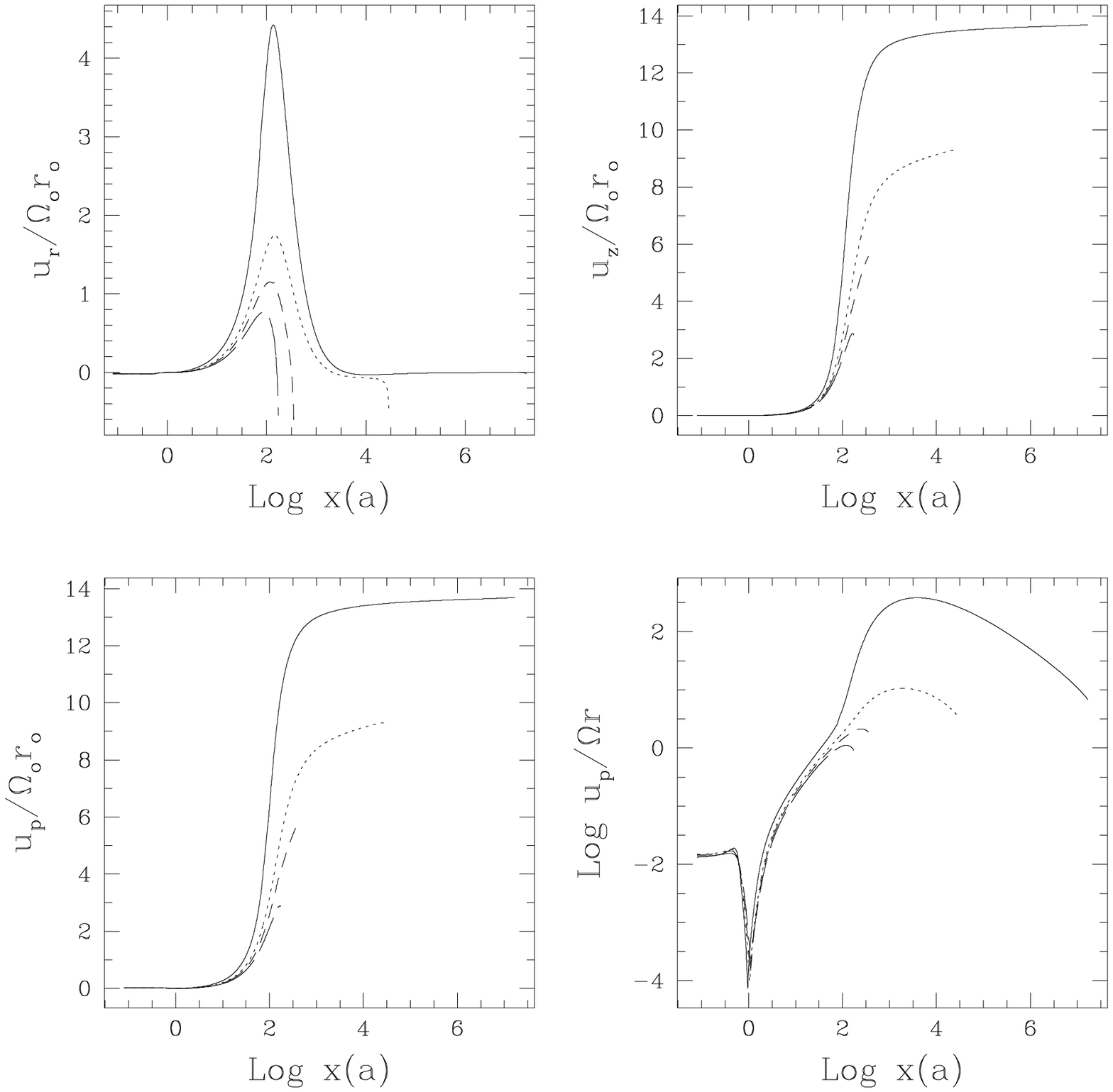,width=7.5cm} &
    \psfig{file=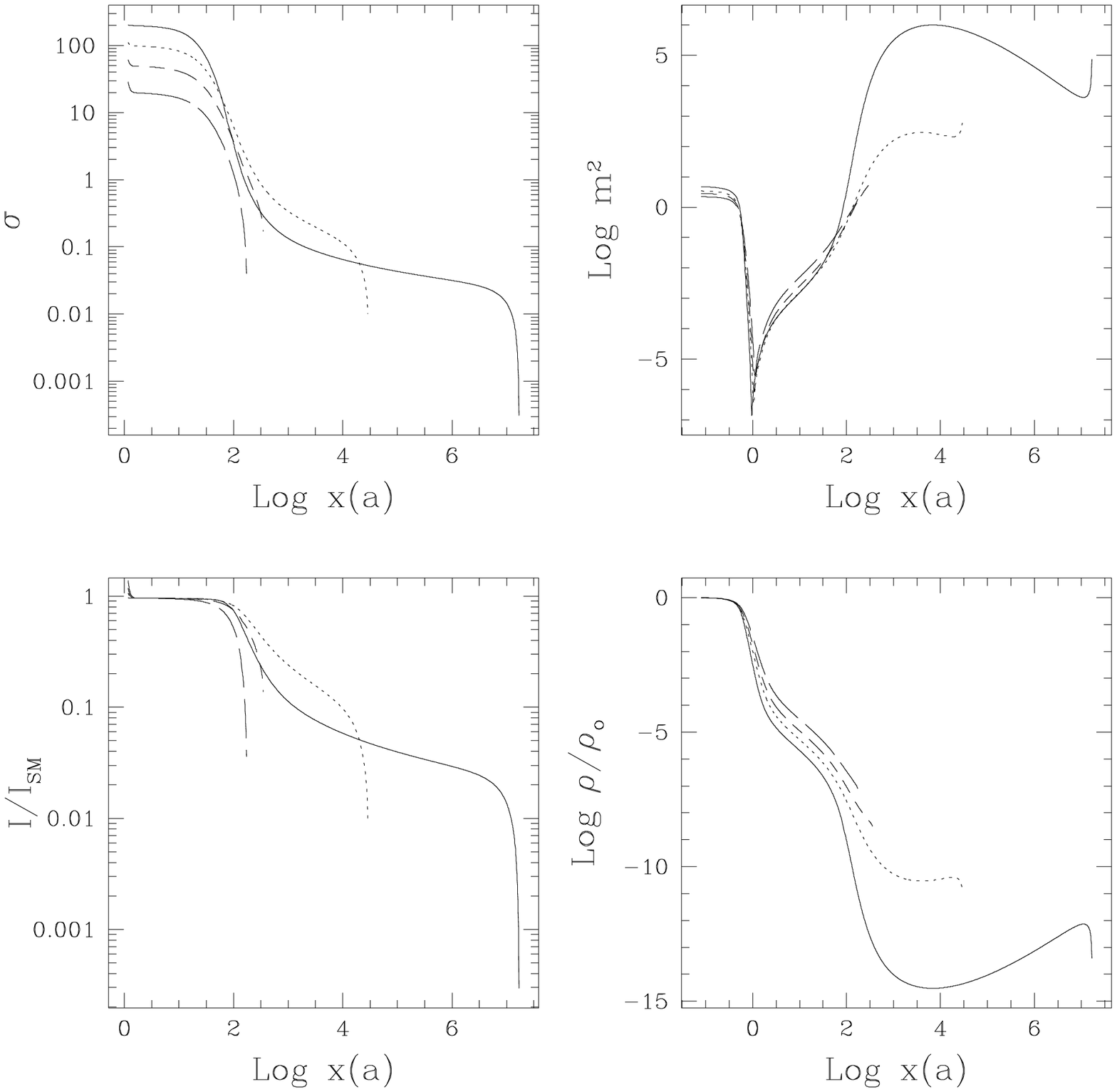,width=7.5cm}\\ 
  \end{tabular}
  \caption[]{Left: Components of the jet poloidal velocity $\vec u_p$ and
    ratio of the poloidal to the toroidal velocities, measured along a
    magnetic surface and different ejection indices: $\xi=0.005$ (solid
    line), $\xi=0.01$ (dotted line), $\xi=0.02$ (short-dashed line) and
    $\xi=0.05$ (long-dashed line). Note that the poloidal velocity is
    almost zero at the disc surface ($x\sim 1$). Right: Ratio $\sigma$ of
    the MHD Poynting flux to the kinetic energy flux, logarithm of the
    Alfv\'enic Mach number $m^2$, total poloidal current $I$ embraced by
    the magnetic surface (normalized to its value at the disc surface) and
    plasma density logarithm along a magnetic surface.}
\end{figure}

Ejected matter leaves the disc with a vertical velocity initially much
smaller than the local sound speed, $u^+_z \simeq m_s \xi C_s$. It gets
however very quickly accelerated as it leaves the resistive MHD zone. The
SM point lies typically between 1 and 2 scale heights, usually at the very
beginning of the ideal MHD zone. Until the Alfv\'en point, plasma is almost
co-rotating with the magnetic field lines, behaving like a rigid
funnel. The Alfv\'en point is far away above the disc, at an altitude $z_A
\sim r_A= \lambda^{1/2} r_o \gg h(r_o)$. After its crossing, there is a
sudden opening of the magnetic surfaces. This is due to the centrifugal
force which is now enhanced by the tension provided by the super-Alfv\'enic
flow ($m^2 >1$, see Eq.~\ref{eq:fperp}). This opening of the magnetic
surfaces is controlled by the quantity $\omega_A$: the larger $\omega_A$,
the larger maximum jet radius. Or, the less current used in the sub-A
region and the more remains in the super-A region.

As the magnetic surface opens up, plasma drags along the field lines which
produces a large toroidal field. This is the consequence of ideal MHD since
matter is rotating {\bf slower} than the field lines ($\Omega <\Omega_*$). 
As the toroidal field increases, the ``hoop stress'' becomes more and more
important until it overcomes the centrifugal force. This is unavoidable if
the jet can open up freely in space (negligible outer pressure) and if its
inner transverse equilibrium allow it: the centrifugal effect decreases
faster with radius than the hoop-stress. As a result, the cold jet
asymptotic transverse equilibrium is simply given by 
\be
- \nabla_{\perp}\left (P_{ext} + \frac{B^2}{2\mu_o}\right) =
\frac{B^2_{\phi}} {\mu_o r} \nabla_{\perp}r
\ee
where $P_{ext}$ must be understood here as the pressure of the material located
on the axis or outside the jet. A strict cylindrical collimation requires
therefore a perfect (and fragile) matching between the hoop-stress and the
total pressure gradient. In our self-similar solutions, $P_{ext}=0$ and the
inner magnetic field gradient is not enough to balance the hoop-stress. As a
consequence, all solutions recollimate (refocus) towards the jet axis. 

This seems to be a (quite) general result of MHD jets that can freely expand
in space. For example, all Blandford \& Payne disc wind solutions and
X-wind solutions obtained by Shu et al. (1995) also recollimate\footnote{Shu
  et al. imposed however a cylindrical asymptotic collimation by assuming
  an equilibrium between the jet hoop-stress and an inner magnetic
  pressure, provided by a poloidal field located on the axis.}. However,
Contopoulos \& Lovelace (1994) as well as Ostriker (1997) obtained
solutions within the same self-similar ansatz that did not recollimate:
recollimation is therefore not a feature of self-similarity alone. On the
other hand, Pelletier \& Pudritz (1992) found recollimating solutions that
are not self-similar. In fact, it is possible to show that if the
conditions
\begin{eqnarray}
\frac{d \ln \lambda}{d\ln r_o} > 1 &\ \ \ \mbox{and}\ \ \ & \frac{d \ln
  (\rho_A/\rho_o)} {d\ln r_o} < \frac{d \ln \rho_o}{d\ln r_o} =\alpha_\rho
\end{eqnarray}
are satisfied in a cold, disc-driven jet, then recollimation would have
been impossible (Ferreira 1997). Evidently, such conditions are violated in
self-similar solutions: $\lambda$ and $\rho_A/\rho_o$ remain constant
throughout the jet. However, this analytical analysis shows that non
self-similar jets produced from a large range of disc radii and displaying
no strong gradients of these quantities may be prone to recollimation (as
in Pelletier \& Pudritz).

\begin{figure}
  \begin{tabular}{cc}
    \psfig{file=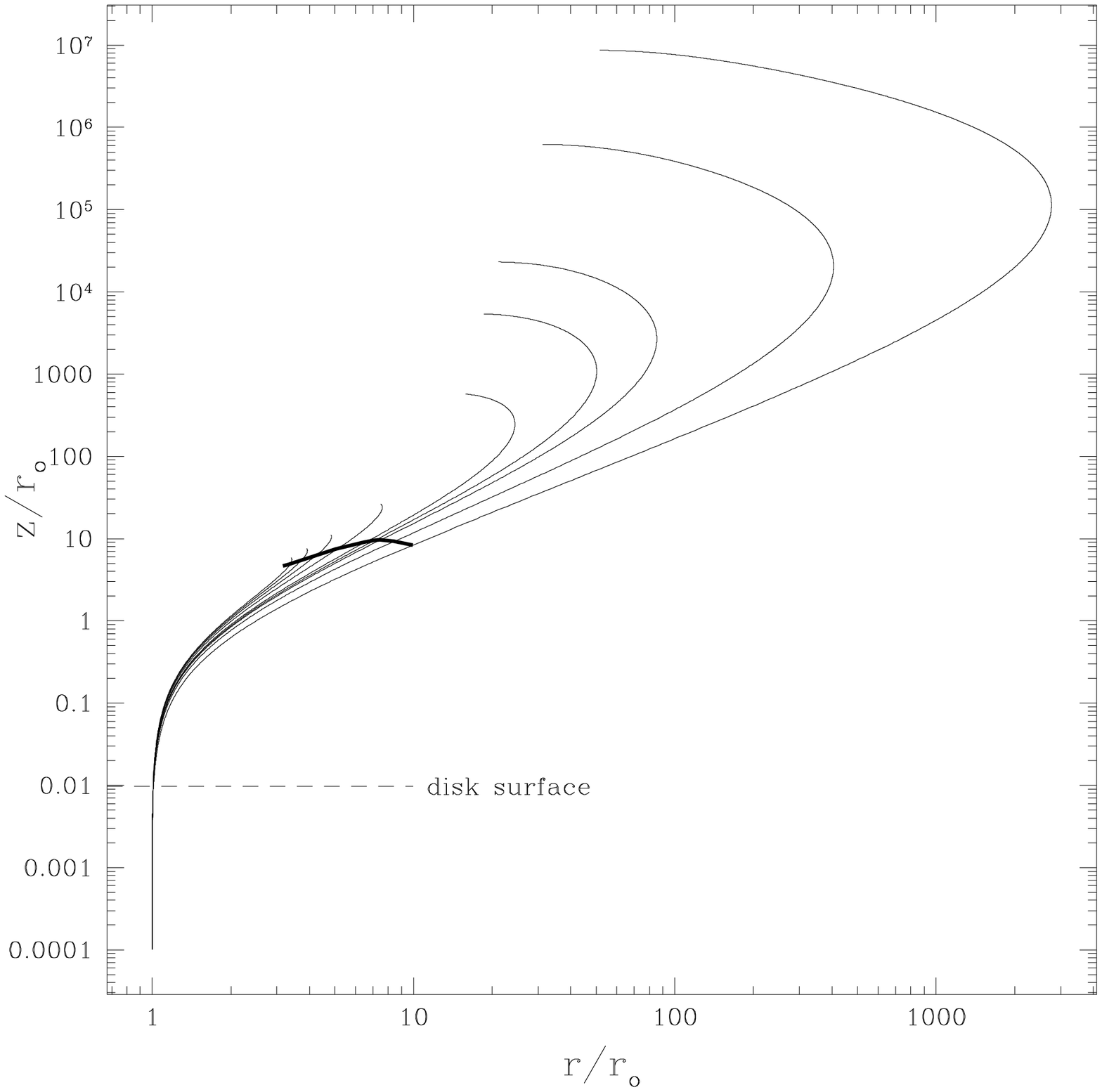,width=8cm} &
    \psfig{file=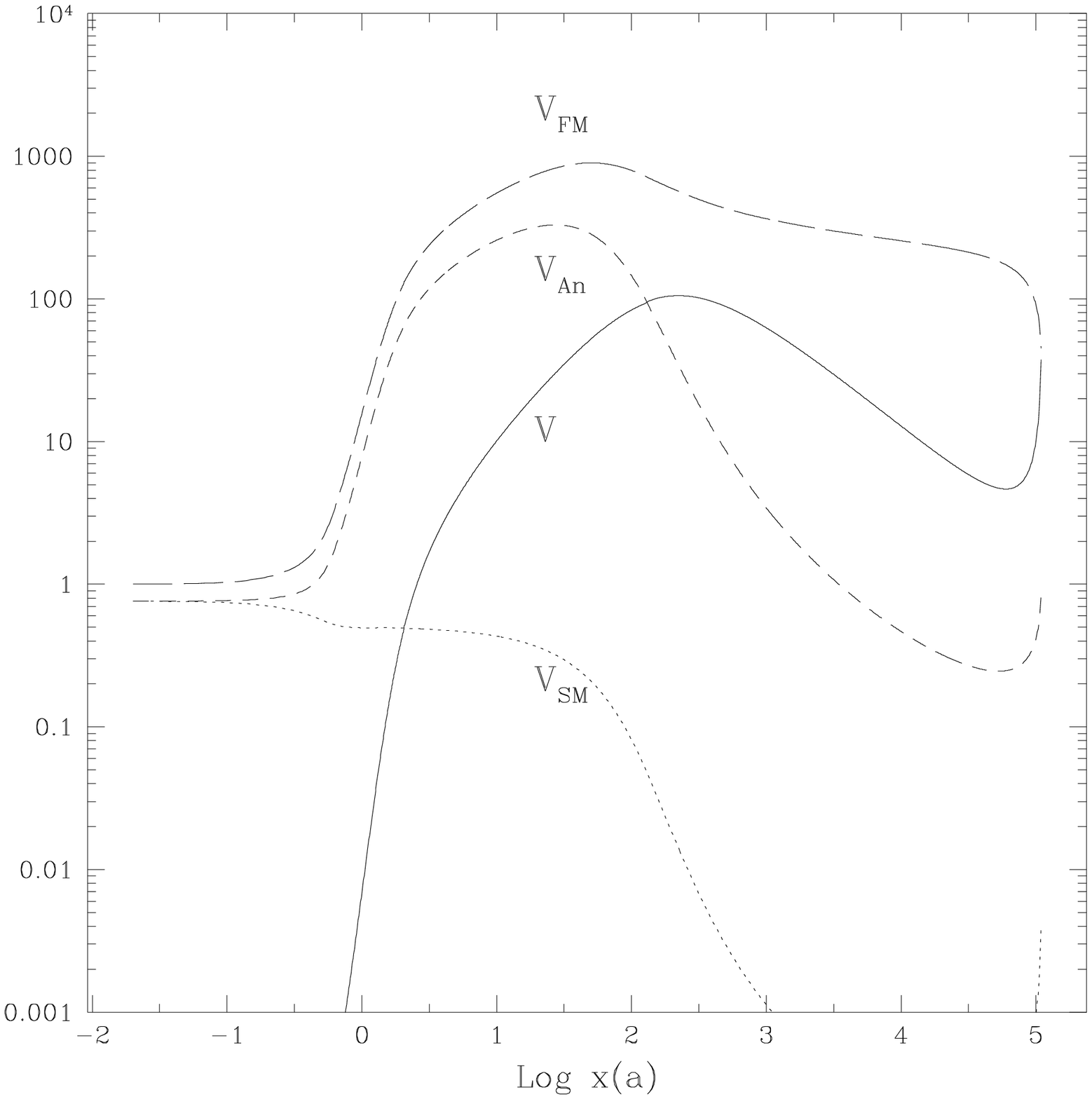,width=8cm}\\ 
  \end{tabular}
  \caption[]{Left: Poloidal magnetic surfaces for $\varepsilon=0.01$ and
    $\alpha_m=1$ and different ejection indices (hence $\omega_A$): $\xi=
    0.05$, 0.04, 0.03, 0.02, 0.012, 0.01, 0.009, 0.007 and 0.005 (the
    maximum radius increases with decreasing $\xi$).  The thick line
    connects the Alfv\'en points of each surface, anchored at a radius
    $r_o$. Right: Critical speed $V= \vec u \cdot \vec n$ and phase speeds
    $V_{SM}$, $V_{An}$, $V_{FM}$ (corresponding respectively to the SM,
    Alfv\'en and FM waves) along a magnetic surface. All speeds are
    normalized to the disc sound speed $\Omega_K h_o$. The SM point lies
    just above the disc surface at $z_{SM} \sim h$, while the Alfv\'en
    point is at $z_A \sim r_A$ (here $\sim 10 r_o$). The solution shown here
    does not cross the last (FM) critical point (although $u_p > V_{FM}$).}
\end{figure}

What happens to our solutions after they recollimate~? When the jet
transverse equilibrium enforces recollimation, almost all available angular
momentum has already been transfered to matter and $\Omega \simeq \Omega_*
r_A^2/r^2$. This implies that the current also goes to zero $I \rightarrow
0$ (and not a constant) and so does the toroidal field. The recollimating
matter drags the field lines along with it, severely reducing the pitch of
the magnetic helix. Before $B_\phi$ reaches zero, our self-similar
solution meets the last critical point, namely $V\sim u_r = V_{FM} \sim
V_{At}$ (note that $u_p > V_{FM}$ already before recollimation). 
Vlahakis et al. (2000) obtained recently super-FM solutions by playing with
the location of the Alfv\'en surface and the jet polytropic index. But these
solutions are terminated in the same way as those displayed here. Such a
behaviour remains unexplained and may be due to the self-similar form of the
solutions. Anyway, such super-FM solutions could safely produce an oblique
shock\footnote{As suggested by Gomez de Castro \& Pudritz (1993) and
  maybe seen in numerical simulations of Ouyed \& Pudritz (1997).} leading
to a time-dependent readjustment of the whole structure. This will not
alter the underlying steady-state solution, for no signal can propagate
upstream. Although Vlahakis et al. solutions are not connected with the
disc, their work gives an indication that introducing another degree of
freedom (the polytropic index value) might indeed provide trans-FM
solutions. But in any case, at this stage, a shock seems the unavoidable
fate of this mathematical class of solutions.

\begin{figure}
  \begin{tabular}{ccc}
      & \centerline{\psfig{file=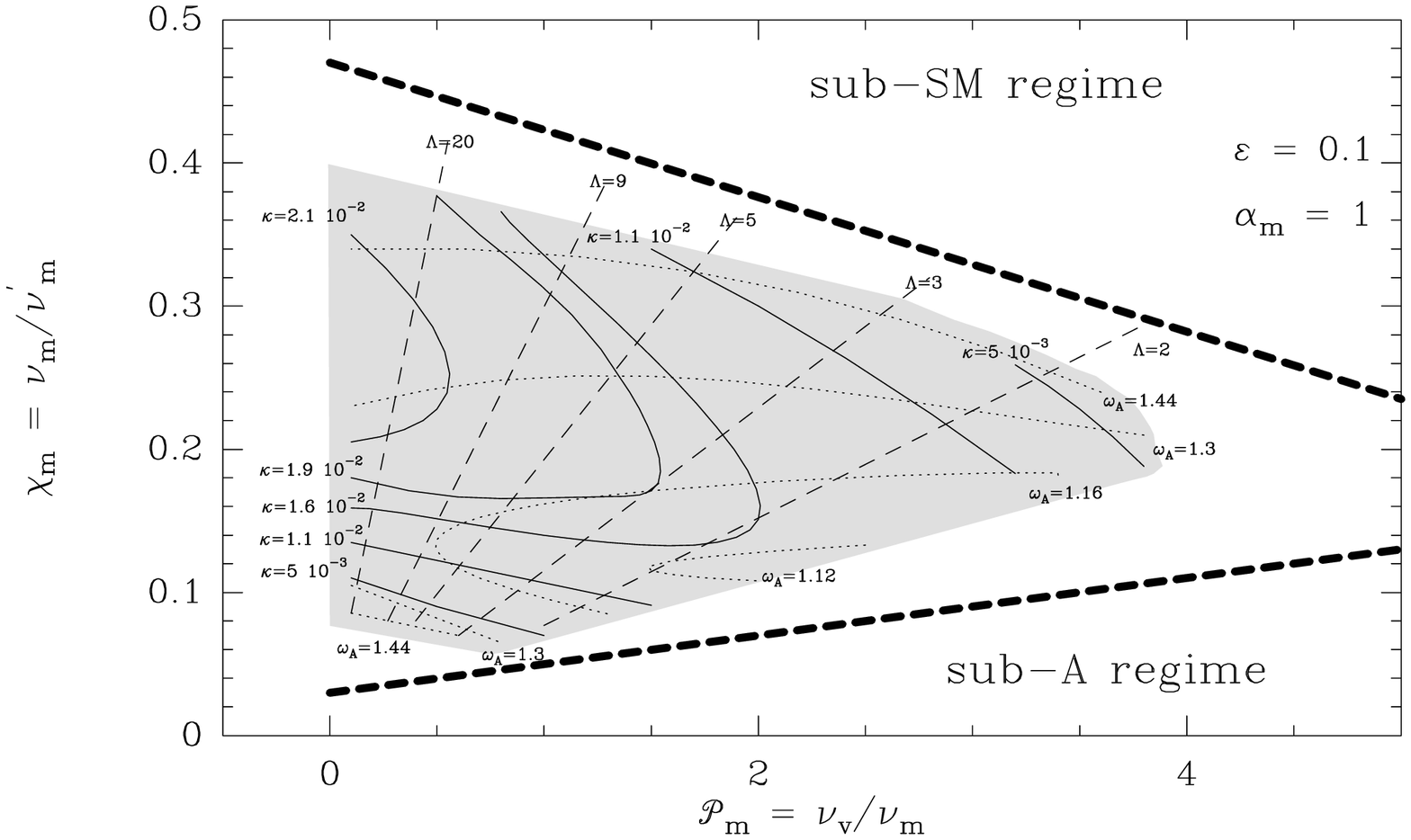,height=5cm,angle=-90}}& \\
     &  \psfig{file=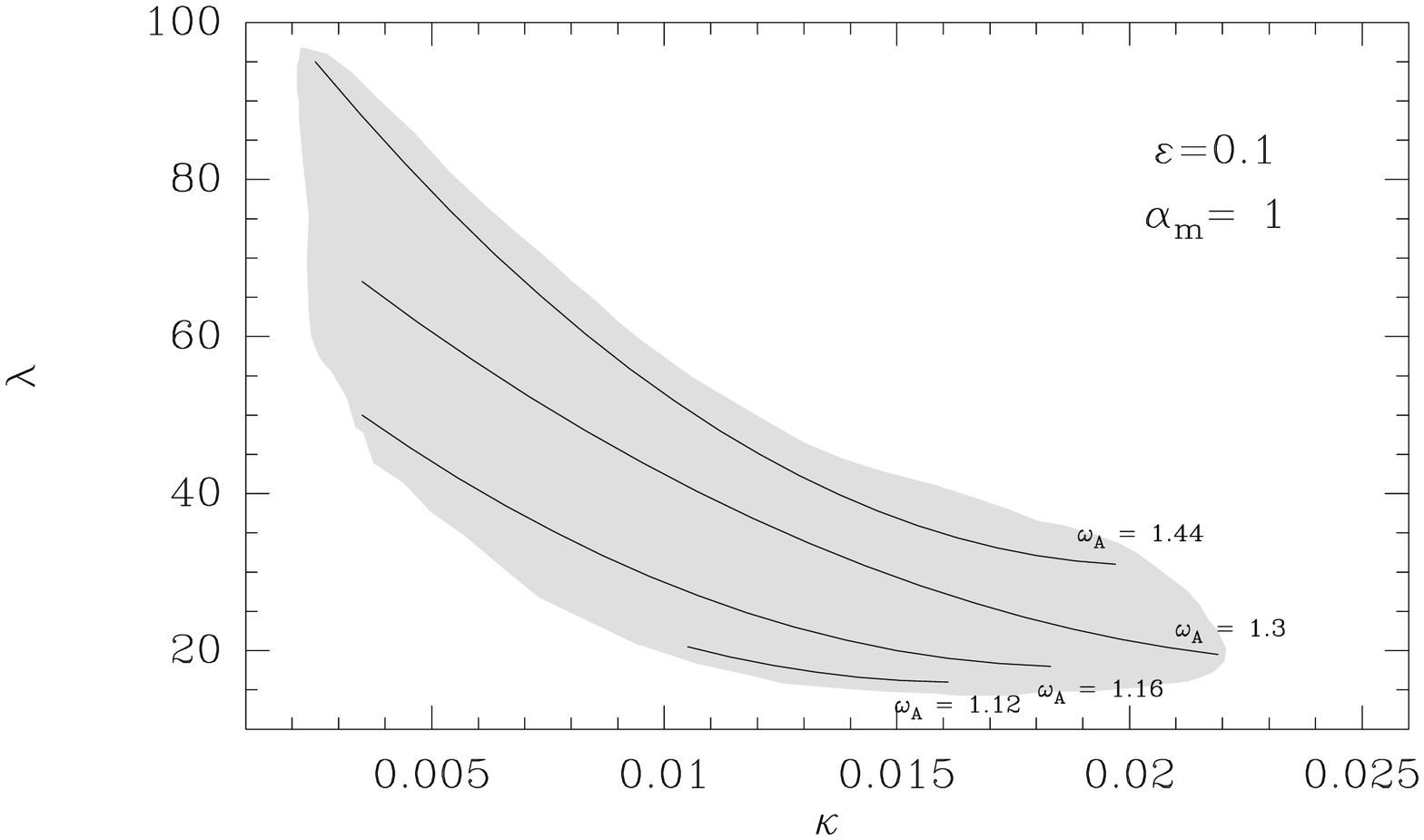,height=5cm,angle=-90}& \\
  \end{tabular}
  \caption[]{Parameter space of cold, adiabatic MAES, for $\alpha_m=1$ and
    $\varepsilon=0.1$. The shaded areas correspond to the location where
    numerical solutions could be found. Top: thick dashed lines show
    theoretical limits, super-SM (upper) and super-A (lower)
    conditions. Levels for the corresponding jet parameters $\kappa$ and
    $\omega_A$ are also displayed. Bottom: the corresponding jet
    $\kappa-\lambda$ parameter space.} 
\label{sp1}
\end{figure}

\subsubsection{Parameter space of cold, adiabatic MAES}

The parameter space of cold MAES is obtained by varying the set of free
disc parameters ($\varepsilon, \alpha_m, {\cal P}_m, \chi_m$). We choose to
fix the values of both $\varepsilon$ and $\alpha_m$ and represent the
parameter space with the remaining parameters (Fig.~\ref{sp1}).
Numerical solutions are only found inside the shaded areas, where we also
plot levels of two jet parameters $\omega_A$ and $\kappa$. This region is
embedded inside a larger region (thick dashed lines), obtained by two
analytical constraints. The first one arises from the requirement
that jets become super-SM and is thus related to the disc vertical
equilibrium. The second emerges from the requirement that jets must become
super-Alfv\'enic, namely $\omega_A> 1$, providing the lower limits in the
same plots. These two analytical constraints strongly depend on both
$\alpha_m$ and $\varepsilon$. Cold, adiabatic MAES have the following
properties: 

$\bullet$ The parameter space is very narrow, with typical values $\xi \sim
10^{-2}$ and $\mu \sim 1$. No solution has been found outside the range
$0.001 < \varepsilon < 0.3$ and $0.3 \leq \alpha_m < 3$. The corresponding
jet parameters lie in the range $10 < \lambda < 100$ and $0.001 < \kappa <
0.03$.

$\bullet$ The parameter space shrinks considerably with $\alpha_m$, because
of the extreme sensibility of $B_\phi^+$ to it (Casse \& Ferreira
2000a). Note however that $\alpha_v= \alpha_m \mu^{1/2} {\cal P}_m$ is
usually smaller than $\alpha_m$.   

$\bullet$ No solution has been found with a dominant viscous torque
($\Lambda \ll 1$): the reason lies in the imposed geometry of the Alfv\'en
surface (conical). On the other hand, high-$\omega_A$ solutions (those with
large jet radius) exist only for magnetically-dominated discs ($\Lambda \gg
1$). 

$\bullet$ The required turbulence anisotropy increases with both $\Lambda$
and $\alpha_m$, following the scaling $\chi_m \sim \Lambda \varepsilon
\alpha_v \alpha_m$ provided by Eq.~(\ref{eq:Ga}).

\subsection{``Warm'' configurations}

\subsubsection{Entropy generation inside the disc}

As seen in Sect.~2.3.2, the disc energy equation (\ref{eq:ener}) is such a 
mess that all works on accretion discs used simpified assumptions. In fact
using an isothermal or adiabatic prescription for the temperature vertical
profile may be an over simplification, that may have led to such a small
parameter space. One way to tackle the energy equation is to solve
\begin{equation}
\rho T\frac{D S}{Dt} =  \rho T \vec u_p \cdot \nabla S = Q
\label{eq:entr}
\end{equation}
where the entropy source $Q= \Gamma - \Lambda$ is prescribed and describes
the {\it local net} effect of all possible heating $\Gamma$ and cooling
$\Lambda$ terms. The sources of heating are:

\hspace{1cm} $\hookrightarrow$ $\Gamma_{eff}= \eta_m J^2_{\phi}\ +
\ \eta'_mJ^2_p\ +\ \eta_v \left| r \nabla \Omega \right|^2$ the effective
Joule and viscous dissipations;  

\hspace{1cm} $\hookrightarrow$ $\Gamma_{turb}$ due to turbulent energy
deposition, not described by anomalous coefficients;

\hspace{1cm} $\hookrightarrow$ $\Gamma_{ext}$ some external source of energy
(like protostellar UV or X-rays, or cosmic rays).

On the other hand, the cooling sources are

\hspace{1cm} $\hookrightarrow$ $\Lambda_{rad}= \nabla \cdot \vec S_{rad}$
radiative losses in optically thick or thin media;

\hspace{1cm} $\hookrightarrow$ $\Lambda_{turb}$ turbulent transport that may be
described by kinetic theory or due to large scale motions like convection
(in Eq.~(\ref{eq:ener}) $\nabla \cdot \vec q_{turb} = \Lambda_{turb} -
\Gamma_{turb}$).    

In our simplified approach, we prescribe both the shape and amplitude
of $Q$, but consistently with energy conservation. We make therefore two
assumptions: (1) there is no net input of turbulent energy in the volume
${\cal V}$ occupied by the MAES, namely $P_{turb}= \int_{\cal V} \left (
\Gamma - \Lambda \right )_{turb} d^3{\cal V}= 0$ and (2) the power
deposited by any external medium is negligible, ie. $P_{ext}= \int_{\cal V}
\Gamma_{ext} d^3{\cal V}= 0$. With these assumptions, the only remaining
source of energy is accretion of the laminar flow: turbulence can only
redistribute energy from one place to another one.

Accretion is possible because of the torque due to the mean magnetic field
(jet) and the turbulent (``viscous'') torque. Accretion energy released by
the first torque is converted into a MHD Poynting flux leaving the disc
(and feeding the jets) and heat through local Joule dissipation. Accretion
energy released by the ``viscous'' torque is conventionaly thought as being
converted into heat through dissipation. But note that if such a torque
arises from field lines connecting two disc radii, one would then expect
also an outward flux of energy, which would be dissipated above the disc
surface (Heyvaerts \& Priest 1989, Miller \& Stone 2000, Machida et
al. 2000). Anyway, these two heating sources ($\Gamma_{eff}$) build up a
local gas thermal energy reservoir which decreases because of the local
cooling terms. Thus, the total power related to dissipation inside the disc
(ie. not directly put into the jets) is $P_{diss}= \int_{\cal V}
\Gamma_{eff} d^3{\cal V}$. In a conventional picture of accretion discs,
such a power is finally radiated away, either at the disc surfaces only, or
also in some chromosphere. Here, we assume that a fraction $f$ of this
power is in fact not lost by the plasma but provides an extra source of
entropy $Q$, namely   
\be
f = \frac{\int_{\cal V} Q d^3{\cal V}}{\int_{\cal V} \Gamma_{eff} d^3{\cal
    V}}  
\ee 
This expression is consistent with global energy conservation, the
parameter $f$ being free and varying from $0$ (``cold'' MAES) to 1
(``warm'' or magneto-thermally driven jets). A value of $f$ larger than
unity would require an extra source of energy. We need now to specify the
vertical (self-similar) profile of $Q$. Obviously, this introduces so many
degrees of freedom that we do not dare anymore to look for the parameter
space. Instead, we will look for extreme configurations and try to derive
quantitative results. 

\begin{figure}[t]
  \begin{tabular}{cc}
    \psfig{file=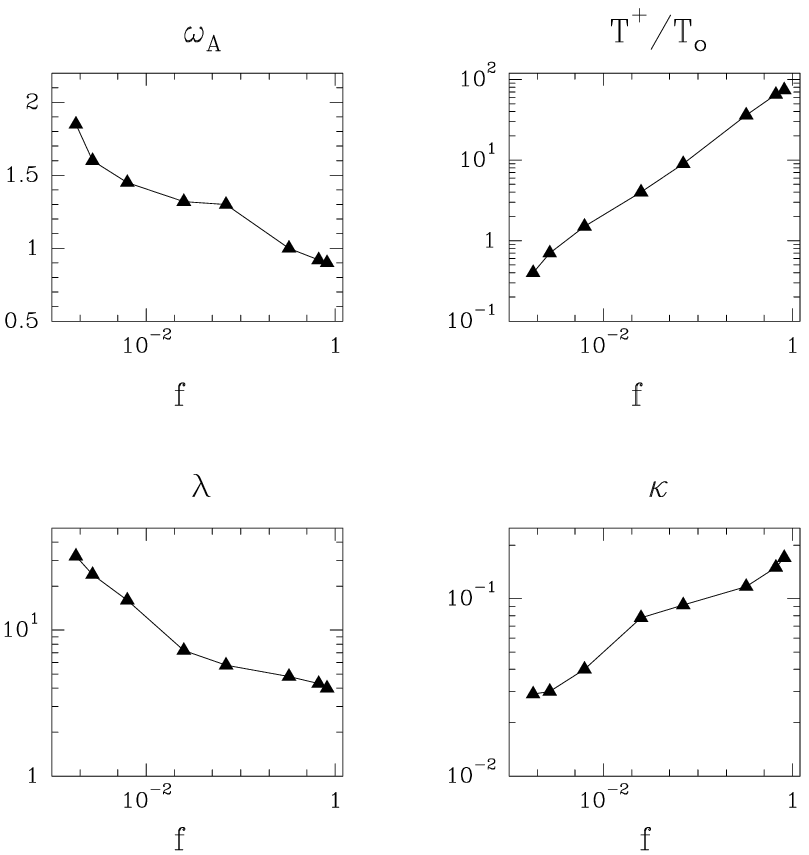,width=8cm} & 
    \psfig{file=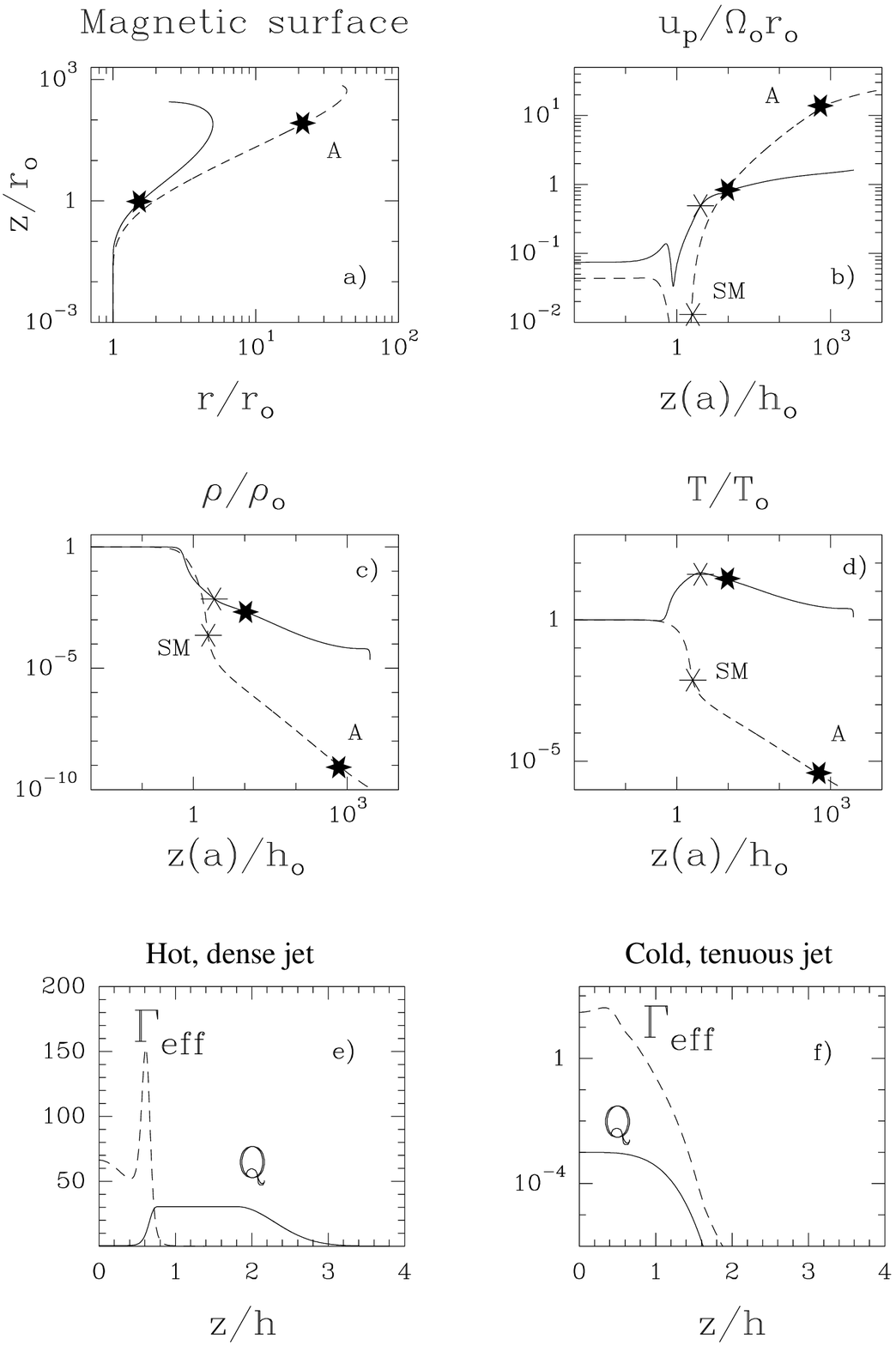,width=8cm}\\
  \end{tabular}
\caption{Left: Variation of jet parameters with $f$. As $f$ increases, jets
  get hotter, denser and with a smaller magnetic lever arm $\lambda$. After
  some threshold, depending on the other MAES (turbulence) parameters,
  thermal effects become so important that super-A jets can be obtained
  with slow-rotators ($\omega_A <1$). Right: Two new extreme MAES with an
  additional heating Q. If Q is very large at the disc surface ($f \sim
  1$), hot and dense jets are produced (solid curves) whereas cold and very
  tenuous jets (dashed curves) are obtained if Q is almost inexistent
  ($f=5\ 10^{-5}$). Lower pannels show the effective turbulent heating
  $\Gamma_{eff}$ and the prescribed entropy source $Q$ at a radius $r_o$,
  normalized to the same quantity.}
\end{figure}

\subsubsection{Dynamical effects}

The steady-state energy equation (\ref{eq:entr}) can  be explicitely written
\be
Q  = \rho T \frac{D S}{Dt}=\rho\frac{D H}{Dt} - \frac{D P}{Dt} = 
\frac{\gamma}{\gamma -1}\frac{k_B}{m_p}\rho\vec u_p \cdot \nabla T -
\vec u_p \cdot \nabla P \ . 
\ee 
We then see that the main influence of a non-zero $Q$ is on the
temperature and pressure vertical gradients (thin disc), thus at the disc
surface. But this is precisely the place where such gradients are
required for ejection (Sect.~4.1.3)! Therefore, allowing some energy
deposition at the disc surface has two major effects: 
\begin{itemize}
\item (1) The initial jet temperature ($T^+$) is increased and may
  thereby provide a non-negligible initial enthalpy $H$ (``warm''
  jets). Indeed, the Bernoulli integral becomes
  \be
  E(a) = \frac{u^2}{2}\ +\ H\ +\ \Phi_G\ -\ \Omega_*\frac{rB_{\phi}}{\eta}\
  -\ \int^s_{s^+} \frac{Q(s',a)}{\rho(s',a)u_p(s',a)}ds' 
  \ee
  where $s$ is a curvilinear coordinate along the magnetic surface and
  $s^+$ represents the disc surface. If $Q$ remains positive above
  $s^+$, it offers an additional energy reservoir for plasma. Moreover, the
  total energy feeding a magnetic surface,
  \begin{equation}
    E(a) = \frac{\Omega_o^2r_o^2}{2}\left(2 \lambda - 3 + \frac{2\gamma}
      {\gamma -1}\frac{T^{+}}{T_{o}} \varepsilon^2 \right) \ . 
  \end{equation}
  is affected by the heating that already occurred in the underlying
  layers. Such an heating may provide a ratio $T^+/T_o$ larger than unity,
  possibly allowing to relax the constraint on minimum field lines
  inclination. 
\item (2) If $Q$ is relevant in the upper resistive layers, then it will
  increase also the plasma pressure gradient. This will enhance the ejected
  mass flux (and lower the magnetic lever arm) and might therefore have
  tremendous consequences on jet dynamics. 
\end{itemize}

We used only one type of vertical profile $f_Q(s)$, changing the value of
the parameter $f$. The chosen profile provides $Q \simeq 0$ inside the disc
($x < 0.5$), an increase until a maximum value (fixed by $f$) around $x
\leq 1$, then a decrease to zero (adiabatic behaviour) after roughly one
scale height (see Casse \& Ferreira 2000b). As $f$ increases (all other
MAES parameters being held constant), one goes from cold ($T^+ < T_o$),
tenuous ($\kappa < 0.02$) jets from fast rotators ($\omega_A >1$) to hot
($T^+ \gg T_o$), dense ($\kappa > 0.1$) jets from slow rotators ($\omega_A
<1$). Namely, the presence of some chromospheric heating allows a smooth
transition from ``cold'' (purely magnetically-driven) to ``warm''
(magneto-thermaly driven) jets.

Another class of ``cold'' solutions can also be designed. Indeed, if local
cooling is not sufficient inside the disc ($Q>0$ for $x<1$), the plasma
pressure increases which provides the disc a stronger support against both
tidal and magnetic compression. As a result, the magnetic field lines can
be more bent than in the previous adiabatic or isothermal solutions. Such a
large curvature hinders mass to be ejected ($\kappa$ may be as small as
$10^{-4}$ and $\lambda \sim 400$) but a vertical equilibrium can
nevertheless be reached\footnote{Note that relativistic speeds are
  expected if such a MAES is settled around a compact object.}. Thus,
entropy generation inside the disc removes the limits found on the ``cold''
parameter space described earlier.

\subsubsection{Global energy conservation}

We suppose that a MAES is settled around a central object of mass $M$,
between an inner radius $r_i$ and some outer radius $r_e$. At this outer
radius, the structure is fed with an accretion rate $\dot M_{ae}= \dot
M_a(r_e)$. Mass conservation then writes $\dot M_{ae} - 2 \dot M_j = \dot
M_{ai}$ and the fraction of ejected mass is
\be
\frac{2 \dot M_j}{\dot M_{ae}} = 1 - \left ( \frac{r_i}{r_e}\right )^\xi
\simeq \xi \ln \frac{r_e}{r_i}
\ee 
the last expression being valid only for a very small ejection efficiency
$\xi \ll 1$. The local energy conservation equation is 
\be
\nabla \cdot \left [ \rho\vec u_p \left ( \frac{u^2}{2}\ +\ \Phi_G\ +\ H
  \right )\ +\ \vec S_{MHD} \ -\  \vec u \cdot {\mathsf T} \right ] = 
\rho T \vec u_p \cdot \nabla S\  -\  \Gamma_{eff}
\label{Econs}
\ee
whereas the second law of thermodynamics (\ref{eq:entr}) provides 
\be
Q= \rho T\vec u_p \cdot \nabla S = \Gamma_{eff}\ +\ (\Gamma_{turb}\
-\ \Lambda_{turb})\ +\ \Gamma_{ext}\ -\  \nabla \cdot \vec S_{rad}
\ee

\begin{figure}[h]
\centerline{\hbox{\psfig{file=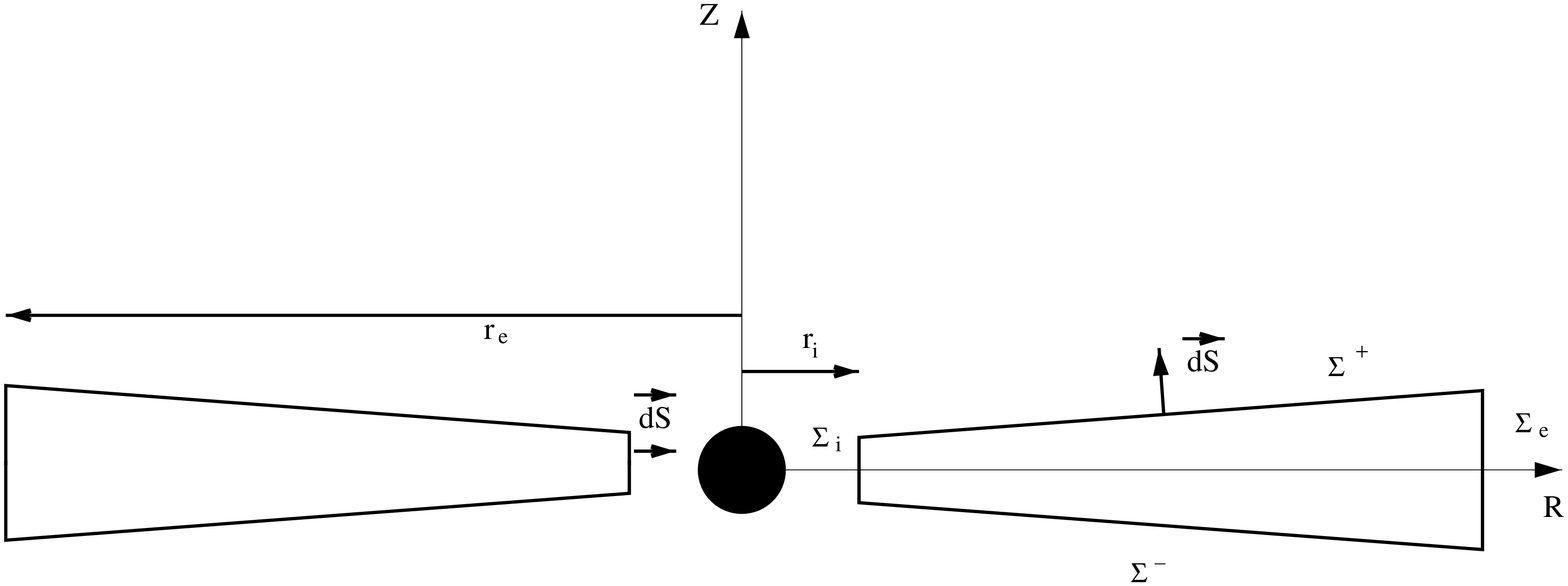,width=14cm}}}
\caption[]{Sketch of the volume used for the calculation of the
  global energy conservation.}
\label{contour}
\end{figure}

To get the global energy conservation, we integrate this equation on the
volume ${\cal V}$ occupied by the disc. We thus define $\Sigma^+$ and
$\Sigma^-$ as the disc surfaces\footnote{The disc surface $z=z^+=x^+h(r)$
  is precisely defined as the locus where $u_r(r,z^+) =0$.} and $\Sigma_i$
($\Sigma_e$) the lateral surfaces at $r=r_i$ ($r=r_e$, see
Fig.~(\ref{contour})). After integration, we get  
\be
P_{acc} + P_{ext} + P_{turb} = 2P_{jet} + 2P_{rad}
\ee
where the accretion power $P_{acc}$, ie. the power released by the
accretion flow, is the difference between what comes in at $r_e$ and goes out
at $r_i$. As said previously, we assumed $P_{ext}=P_{turb}= 0$ (neither an
external source of energy, nor a significant input of turbulent
energy). Thus, all available power $P_{acc}$ is shared between radiative
losses at the disc surfaces $P_{rad}= \int_{\Sigma^{\pm}} \vec{S}_{rad}
\cdot \vec{dS}$ and jet power $P_{jet}= \int_{\Sigma^{\pm}} \rho \vec{u_p}
E(a) \cdot \vec{dS}$. It is usefull to introduce the {\it fiducial}
quantity
\be
P_{lib} \equiv  \eta_{lib} \frac{G M \dot{M}_{ae}}{2 r_i} \simeq 2.5\,
10^{32}\ \left (\frac{M}{M_\odot}\right ) \left (\frac{\dot M_{ae}}
  {10^{-7} M_\odot/yr} \right ) \left (\frac{r_i}{0.1 \mbox{ AU}}\right
)^{-1}\  \mbox{erg s}^{-1} 
\ee
where $\eta_{lib}$ is a term roughly equal to unity. Within our
self-similar framework, energy conservation of a thin (or slim) disc writes
\begin{eqnarray}
\frac{P_{acc}}{P_{lib}} & = & (1 - \xi)\left (1 + \frac{1}{2}\frac{\varepsilon
  \Lambda}{1+\Lambda} \right )\\
\frac{2P_{jet}}{P_{lib}} &=& \frac{\Lambda}{1+\Lambda}\left|
  \frac{B^+_{\phi}}{qB_o}\right|\  +\  \frac{2\gamma}{\gamma-1}
\frac{T^+}{T_o}\xi\varepsilon^2 \ -\  \xi\\
\frac{2P_{rad}}{P_{lib}} &=& (1 - f) \frac{P_{diss}}{P_{lib}} =
\frac{P_{acc}-2P_{jet}}{P_{lib}}
\end{eqnarray} 

See Casse \& Ferreira (2000b) for the derivation of these
expressions. Three important remarks. First, the absolute limit for the 
ejection efficiency is $\xi= 1$; Second, the real energy release is
$P_{acc}$, comparable to  $P_{lib}$ only for low ejection indices. Finally,
the MHD Poynting flux feeding the jets depends directly on the amount of
toroidal field at the disc surface. Thus, magnetically-dominated discs
($\Lambda \gg 1$) may still produce some disc luminosity, provided $|
B^+_{\phi}| < qB_o$ (and $f < 1$). For $f \sim 0$ we found solutions with a
ratio $P_{jet}/P_{rad}$ varying between 0.1 and 10.

\subsubsection{Preliminary conclusions}

From this study, we are forced to conclude that thermal effects have an
outrageous quantitative importance on jet launching. In a way, this is
fortunate since we can now recover and understand the parameter space
obtained with numerical MHD simulations. Although ``cold'', the simulated
jets have enormous mass loads and correspondingly small magnetic lever arms
((eg. $\kappa \sim 0.6$, $\lambda \sim 3.5$, Ouyed \& Pudritz 1999). This
may be easily achieved from an accretion disc where a significant entropy
generation took place at the disc surface ($f>0.1$), followed right
afterwards by a strong cooling ($Q<0$ for $x>1$). In this way, very dense
but cold (well, after some time) jets can indeed be produced. Remember that
the bottom of the computational box has nothing to do with the real disc
surface. 

A major drawback of such a conclusion is that any quantitative prediction
(eg. the value of $\xi$) requires to treat the energy equation. We must
therefore inject our knowledge on MHD turbulence and its (possibly non
local) energy transport properties~! Besides, illumination effects by the
central protostar may also be important (since it heats up the disc surface).

\subsection{Observational predictions}

\subsubsection{Accretion discs}

Whether or not the disc remains geometrically thin depends on its internal
temperature, hence on the energy equation. Although the energy transport
processes are unknown, an opaque disc must radiate at its surfaces all
deposited energy. This translates into an effective temperature at the
photosphere $z=h_{phot}(r)$
\be
\sigma T_{eff}^4 = \int_0^{h_{phot}} \nabla \cdot \vec S_{rad} \, dz =
(1-f) \int_0^{h_{phot}} \left [ \eta_m J^2_{\phi}\ +\ \eta'_mJ^2_p\ +\ \eta_v
\left| r \nabla \Omega \right|^2 \right ]\, dz 
\ee
The rhs of this equation can be estimated whereas the effective and central
temperatures are crudely related through $T_{eff}\sim T_o \tau^{-1/4}$, 
where $\tau \sim \kappa \rho_o h_{phot}$ is the disc optical depth. Now,
assuming that $h_{phot}(r) \sim h(r)$ and choosing the dominant disc
opacity to be the grain opacity $\kappa= 0.1\ T^{1/2}\ \mbox{cm}^2
\mbox{g}^{-1}$ (Bell \& Lin 1994), allow us to estimate both the local disc
aspect ratio $\varepsilon$ and optical depth $\tau$ without even solving
the energy equation. Once we have $\varepsilon= h/r$, we can express all disc
quantities in terms of the remaining free parameters. Hence, steady-state
theory of MAES requires the following physical conditions in a disc settled
around a low-mass protostar: 
\begin{eqnarray*}
  \frac{h}{r} &= & 5\ 10^{-2}\ \alpha_v^{-1/9} (1-f)^{1/9}
  \left (\frac{M}{M_\odot}\right )^{-1/3} 
  \left (\frac{\dot M_{eff}}{10^{-7}M_\odot/yr}\right )^{2/9} \\ 
  \tau &\simeq & 7.5\ \alpha_v^{-8/9} (1-f)^{-1/9} \left (\frac{M}{M_\odot}
  \right )^{1/3} \left (\frac{\dot M_{eff}}{10^{-7}M_\odot/yr}\right )^{7/9} 
  \left (\frac{r_o}{1\ \mbox{AU}}\right )^{-1} \\ 
  T_{eff} &= & 168\ (1-f)^{1/4} \left (\frac{M}{M_\odot}\right )^{1/4} 
  \left (\frac{\dot M_{eff}}{10^{-7}M_\odot/yr}\right )^{1/4} 
  \left (\frac{r_o}{1\ \mbox{AU}}\right )^{-3/4} \mbox{ K}\\ 
  T_o &=& 276\ \alpha_v^{-2/9} (1-f)^{2/9} \left (\frac{M}{M_\odot}\right 
  )^{1/3} \left (\frac{\dot M_{eff}}{10^{-7}M_\odot/yr}\right )^{4/9} 
  \left (\frac{r_o}{1\ \mbox{AU}}\right )^{-1} \mbox{ K}\\ 
  \Sigma_o=2\rho_o h &=& 9\ \alpha_v^{-7/9} (1-f)^{-2/9} \left
    (\frac{M}{M_\odot} \right )^{1/6} \left (\frac{\dot
      M_{eff}}{10^{-7}M_\odot/yr}\right )^{5/9}  
  \left (\frac{r_o}{1\ \mbox{AU}}\right )^{-1/2} \mbox{ g.cm}^{-2}\\ 
  B_o &=& 1.3\ \mu^{1/2} \alpha_v^{-4/9} (1-f)^{-1/18} \left 
    (\frac{M}{M_\odot}\right )^{5/12}\left (\frac{\dot M_{eff}}{10^{-7}
      M_\odot/yr}\right )^{7/18} \left (\frac{r_o}{1\ \mbox{AU}}\right 
  )^{-5/4} \mbox{ G}\\
  \frac{u_o}{\Omega_K r_o} &=&  3\ 10^{-3}\ \alpha_v^{7/9} (1-f)^{2/9}
  (1+\Lambda)^{5/9} \left (\frac{M}{M_\odot}\right )^{-2/3} 
  \left (\frac{\dot M_a(r_o)}{10^{-7}M_\odot/yr}\right )^{4/9}
\end{eqnarray*}
where $\dot M_{eff}= \dot M_a(r_o)/(1 + \Lambda)$. Parameters $\mu \sim 1$ and
$\xi$ are provided by the MHD solution. Note that the self-similar scaling
$h(r) \propto r$ is here slightly violated since $\dot M_a \propto
r_o^\xi$. But such a deviation remains minor for $\xi < 0.1$. Care must be
taken when looking at those expressions because $\Lambda$ is imposed by
Eq.~(\ref{eq:Ga}) and depends thus on ($\varepsilon, \alpha_m, {\cal P}_m,
\chi_m$). But this choice allows to recover both radial scalings and values
of a standard accretion disc when $\xi=f=\Lambda=0$. To get a super-A
solution with $\alpha_v= \alpha_m {\cal P}_m \mu^{1/2} \ll 1$ one probably
needs to play around with $f$ (for $f=0$ the parameter space is already
known and requires large values of $\alpha_v$).  

Several observational diagnostics could reveal the presence of a magnetized
disc driving jets: (i) the measure of a large (organized) disc magnetic
field; (ii) optically thin regions or some lack of disc emission (for models
with $\Lambda \gg 1$); (iii) a different spectral energy distribution (the
effective temperature scales as $r_o^{(\xi - 3)/4}$) for discs with a large
ejection index $\xi$. But usually, SED are tricky to interpret and we
don't have the resolution yet to measure disc magnetic fields.

\subsubsection{Self-collimated jets}

The most acurate tool to discriminate between models is to confront
theoretical predictions with recent {\it spatially resolved} observations
of the inner wind structure of TTS in forbidden emission lines of
[O~{\sc i}], [S~{\sc ii}] and [N~{\sc ii}]. Indeed, being optically thin,
these lines carry information on both dynamic and thermodynamic properties
of the whole volume of emission. So, one way to use this information is to
construct the following synthetic observations and comparing them to real
ones: 
\begin{itemize}
\item {\bf Emission maps}, which must then be convolved to typical
  resolutions in order to predict what would be the observed jet morphology
  and collimation properties (eg. displacement of emission peaks, jet FWHM
  as a function of distance).  
\item {\bf Line profiles}, like those obtained using long-slit
  spectroscopy, and integrated profiles. 
\item {\bf Integrated line fluxes} as well as their correlations with the
  disc accretion rate.
\item {\bf Forbidden line ratios}, which reflect the values of the electron
  density, ionisation fraction and local temperature.  
\end{itemize}
While the first two observations offer constraints on the jet dynamics, the
last ones test mainly the heating and ionization mechanisms along the jet. 
We already have a dynamical model providing us with jet density and
velocity. The gas emissivity then requires to know the jet thermal and 
ionization states. With that in hand, we can easily compute synthetic
observations and compare them to real ones. This has first been done by
Safier (1993) using Blandford \& Payne jet solutions. We can now do the
same with models of MAES that take self-consistently into account the
disc-jet interrelations.  

Cabrit et al. (1999), using parameterized temperature and ionization
fraction, produced synthetic maps and long-slit spectra that were nicely
compatible with observations (Shang et al. (1998) did the same for
X-winds). But more reliable predictions require actually solving for the
jet thermal and ionization state, given some local heating mechanism. Our
models being stationary, we cannot invoke shocks as being the heating
process. Moreover, this would require the introduction of additional free 
parameters~! This is even worse if we rely on some small scale turbulence,
providing a means to dissipate locally jet kinetic energy. The only
self-consistent mechanism, intrinsic to MHD flows of low ionization, is
ambipolar diffusion (see Sect.~2.2.3). 

Using this (unavoidable) effect to heat up jets, Garcia et al. (2001)
solved the energy equation along the jet as well as its ionization state,
taking into account several heavy elements (C, N, O, S, Ca, Mg, Fe...),
photoionization heating and cooling by Hydrogen recombination lines. They
used cold (isothermal) MAES solutions obtained by Ferreira (1997) with $\xi
\sim 0.01$. Jet widths and variations in line profiles with 
distance and line tracer are well reproduced. However, predicted maximum
velocities are too high, total densities too low and, as a result, the
low-velocity [O~{\sc i}] component is too weak. Thus, denser and slower MHD
jets are required.

In the solutions used, ejected plasma reaches its asymptotic speed
$u_\infty \simeq \Omega_o r_o \sqrt{2 \lambda - 3}$, which is 
typically 10 times the Keplerian speed at the field line footpoint. Even
though there is some inclination effect that decreases the observed jet
velocity, it is still too large: emission from the jet inner region is
important. Thus, one needs to decrease the jet terminal speed, which
requires to diminish the magnetic lever arm $\lambda$, ie. to increase the
ejection index $\xi$ (which automatically provides a denser jet). However,
models with values of $\xi$ larger than 0.01 require an additional heating
at the disc surface. How much exactly remains to be worked out. 

\section{Some perspectives}

At this stage, one can have confidence in several things: (1) we know
exactly, in a model independent way, how accretion discs can steadily drive
jets; (2) jet properties (velocity, collimation) are strongly dependent on
the mass flux allowed by the disc; (3) such a mass flux is highly sensitive
to the critical disc energy equation; (4) we don't know if real accretion
discs ever meet the required physical conditions for MAES.

The MAES paradigm is based on the belief that a large scale magnetic
field is threading the disc {\it on a large radial extension}. This ensures
some kind of cylindrical geometry (actually a conical Alfv\'en surface),
whereas ejection from a small disc region, as in X-winds (Shu et al. 1994),
would provide a spherical expansion of the field lines (and a more or less
spherical Alfv\'en surface). Such a basic fact provides very different jet
behaviours (see appendix B in Casse \& Ferreira 2000a), but the underlying
accretion/ejection mechanism remains exactly the same. As we saw, comparing
synthetic to real observations is quite a powerful tool to eliminate MAES
models and will probably help to discriminate between ``disc winds'' and
``X-winds''. We know now that, inside the MAES paradigm, YSOs require
active chromospheres and coronae\footnote{Kwan (1997) reached the same
  conclusion from energetic requirements on the low-velocity component of
  emission lines.}. Such a work must therefore be continued, in 
the hope that a characterization of their properties can indeed be
achieved. In parallel, a thorough theoretical (analytical and numerical)
work must be performed to understand how instabilities in magnetized discs
may lead to turbulence and anomalous transport.

What is not addressed by current models of MAES, is the interaction with
the central object. However, including such an interaction and even
allowing for some energetic events to occur there (eg. such as unsteady 
mass ejection) would not perturb the above picture. The main strength of
the simplified version studied here (pure ``disc winds'') is to provide clear
answers to the basic phenomena of correlated accretion and
ejection. Anyway, such an interaction is unavoidable and, for a young 
protostar, may be the ingredient solving the ``angular momentum
problem''. This problem arises from the idea that a protostar of masse $M$
and radius $R$ arises from the collapse of a rotating molecular cloud. The
specific angular momentum of these clouds is typically $10^{21}\mbox{
  cm}^2$/s (Goodman et al. 1993). A typical T-Tauri star (optically
revealed protostar) rotates with a period of 8 days, which provides a
specific angular momentum of the order of some $10^{17}\mbox{
  cm}^2$/s. Where has all this angular momentum gone~? Evidently, we can
assume that a (magnetic) braking occurred during the collapse, but no model
provided yet a good account of the angular momentum transport. A
conservative approach is to assume that, once formed, the protostar is
rotating close to break-up, with a period $P= 2\pi \sqrt{R^3/GM}$ ($\sim$ 1
day). But the corresponding specific angular momentum is still an order of
magnitude larger than that of a T-Tauri star. Thus, another very efficient
braking must occur during the embedded phase. Moreover, there are some
observational evidences that accretion discs play a role in braking down
T-Tauri stars. Indeed, even though they are contracting and accreting
material, T-Tauri stars seem to maintain a constant angular velocity
(Bouvier et al. 1997). The idea is therefore to imagine an interaction with
the accretion disc providing a negative torque on the protostar.

If nothing prevents the accreting disc material to approach the stellar
surface, an equatorial boundary layer will form at the interface
(eg. Popham et al. 1993). But because matter is rotating faster than the
protostar, such an interaction can only lead to a spin-up of the
protostar. The current paradigm is a magnetospheric interaction with the
disc. The protostar is believed to have a large scale magnetic field able
to truncate the disc by forcing the accreting material to leave the
equatorial plane and follow the magnetosphere. A spin-down is then obtained
whenever the field is large enough so that the truncation radius
(magnetopause) is located {\bf beyond} the co-rotation radius $r_{co}$,
defined by $\Omega_{star}= \Omega_{disc}(r_{co})$. In this case, the
protostellar angular momentum is deposited into the inner disc
region... which then must expell it~!

Some models rely on a very efficient disc turbulent viscosity that would
radially transport both disc and protostellar angular momentum (eg. Li
1996). Or, if not throughout all the disc, just on a small region until it
reaches a region located beyond the closed magnetosphere. There, open field
lines could transport this excess of angular momentum in a jet (``X-winds''
by Shu et al. 1994). As said previously, these X-winds carry away the exact
angular momentum at a rate allowing accretion: they are therefore
disc-winds, but produced from a tiny disc region (see also Fendt \& Elstner
2000). It may be more economic to use the protostellar angular momentum and
rotational energy to directly produce another jet component. This can be
done whenever the protostellar magnetic moment has the same polarity as the
disc magnetic field. A magnetic neutral line forms at the equatorial plane,
where reconnection of the two fields takes place. Above this reconnection
site, a fraction of the accreting matter can be loaded onto newly opened
protostellar field lines (Ferreira et al. 2000). Such Reconnection X-winds
arise from a different ejection mechanism: they expell disc material thanks
to the protostellar rotational energy. This is the reason why they can
brake down very efficiently the protostar, on time scales compatible with
observations.

It took almost 30 years to {\it prove} that accretion discs can indeed launch
jets that carry away a sizeable fraction of their angular
momentum. Investigations of the magnetospheric interaction between a
protostar and its circumstellar disc are just beginning and the situation
is far more complex than for discs. Lots of analytical efforts have still
to be done, probably guided by the insights provided by numerical
simulations.


\begin{references}

\reference 
Balbus A.S., Hawley J.F. 1991, ApJ 376, 214\\
A powerful local shear instability in weakly magnetized disks.I- Linear
analysis  

\reference
Bell K.R., Lin D.N.C. 1994, ApJ 427, 987\\ 
Using FU Orionis outbursts to constrain self-regulated protostellar disk
models 

\reference
Blandford R.D., Rees M.J. 1974, MNRAS 169, 395\\
A ``twin-exhaust'' model for double radio sources

\reference
Blandford R.D. 1976, MNRAS 176, 465\\
Accretion disc electrodynamics- A model for double radio sources

\reference 
Blandford R.D., Payne D.G. 1982, MNRAS 199, 883\\
Hydromagnetic flows from accretion discs and the production of radio jets

\reference
Bouvier J., Forestini M., Allain S. 1997, A\&A 326, 1023\\
The angular momentum evolution of low-mass stars.

\reference
Brandenburg A., Donner K.J. 1997, MNRAS 288, L29\\
The dependence of the dynamo alpha on vorticity

\reference
Breitmoser E., Camenzind M. 2000, A\&A 361, 207\\
Collimated outflows of rapidly rotating young stellar objects. Wind
equation, GSS equation and collimation

\reference
Bridle H.A., Perley A.R. 1984 ARA\&A 22, 319\\
Extragalactic radio jets

\reference
Cabrit S., Ferreira J., Raga A.C. 1999, A\&A 343, L61\\
Forbidden lines from T Tauri disk winds. I. High magnetic torque models

\reference 
Camenzind M. 1990, in G. Klare (ed.), Rev. in Modern Astrophysics, 3,
Springer-Verlag, Berlin\\
Magnetized Disk-Winds and the Origin of Bipolar Outflows.

\reference
Cant{\'o} J. 1980, A\&A 86, 327\\
A stellar wind model for Herbig-Haro objects

\reference
Cao X., Jiang D.R. 1999, MNRAS 307, 802\\
Correlation between radio and broad-line emission in radio-loud quasars

\reference
Casse F., Ferreira J. 2000a, A\&A 353, 1115\\
Magnetized accretion-ejection structures. IV. Magnetically-driven jets from
resistive, viscous, Keplerian discs

\reference
Casse F., Ferreira J. 2000b, A\&A 361, 1178\\
Magnetized accretion-ejection structures. V. Effects of entropy generation
inside the disc 

\reference 
Chan K.L., Henriksen R.N. 1980, ApJ 241, 534\\
On the supersonic dynamics of magnetized jets of thermal gas in radio
galaxies 

\reference
Ciolek G.E., Mouschovias, T. 1995, ApJ 454, 194\\
Ambipolar Diffusion, Interstellar Dust, and the Formation of Cloud Cores
and Protostars. IV. Effect of Ultraviolet Ionization and Magnetically
Controlled Infall Rate  

\reference 
Contopoulos J., Lovelace R.V.E. 1994, ApJ 429, 139\\
Magnetically driven jets and winds: Exact solutions

\reference
Contopoulos J., Sauty C. 2001, A\&A 365, 165\\
The origin of molecular protostellar outflows

\reference
Crutcher R.M. 1999, ApJ 520, 706\\
Magnetic Fields in Molecular Clouds: Observations Confront Theory

\reference
DeCampli W.M. 1981, ApJ 244, 124\\
T Tauri winds

\reference 
Fendt C., Camenzind M., Appl S. 1995, A\&A 300, 791\\
On the collimation of stellar magnetospheres to jets. I. Relativistic
force-free 2D equilibrium. 

\reference
Fendt C., Elstner D. 2000, A\&A 363, 208\\
Long-term evolution of a dipole type magnetosphere interacting with an
accretion disk. II. Transition into a quasi-stationary spherically radial
outflow 

\reference 
Ferreira J., Pelletier G. 1993, A\&A 276, 625\\
Magnetized accretion-ejection structures. I. General statements

\reference 
Ferreira J., Pelletier G. 1995, A\&A 295, 807\\
Magnetized accretion-ejection structures. III. Stellar and extragalactic
jets as weakly dissipative disk outflows.

\reference 
Ferreira J. 1997, A\&A 319, 340\\
Magnetically-driven jets from Keplerian accretion discs

\reference
Ferreira J., Pelletier G., Appl S. 2000, MNRAS 312, 387\\
Reconnection X-winds: spin-down of low-mass protostars

\reference
Garcia P., Ferreira J., Cabrit S., Binette L., Raga A., Dougados C., Casse
F., Lavalley C. 2001 in ``Emission Lines from Jet Flows'' Isla Mujeres
(Mexico)\\ 
Forbidden line emission from MHD disk winds and time-variable jets:
comparison with T Tauri microjets

\reference 
Gomez de Castro A.I., Pudritz R.E. 1993, ApJ 409, 748\\
The origin of forbidden line emission from young stellar objects

\reference
Goodman A.A., Benson P.J., Fuller G.A., Myers P.C. 1993, ApJ 406, 528\\
Dense cores in dark clouds. VIII - Velocity gradients

\reference
Hartmann L., McGregor K.B. 1982, ApJ 259, 180\\
Protostellar mass and angular momentum loss

\reference 
Heyvaerts J., Norman C.A. 1989, ApJ 347, 1055\\
The collimation of magnetized winds

\reference
Heyvaerts J., Priest E.R. 1989, A\&A 216, 230\\
A model for a non-Keplerian magnetic accretion disk with a magnetically
heated corona 

\reference
Heyvaerts J., Priest E.R., Bardou A. 1996, ApJ 473, 403\\
Magnetic Field Diffusion in Self-consistently Turbulent Accretion Disks

\reference
Jones D.L., Werhle A.E., Meier D.L., Piner B.G. 2000, ApJ 534, 165\\
The radio jets and accretion disk in NGC 4261

\reference
K\"onigl A. 1982, ApJ 261, 115\\
On the nature of bipolar sources in dense molecular clouds

\reference
K\"onigl A. 1986, Can. J. Phys. 64, 362\\
Stellar and galactic jets: theoretical issues 

\reference
Krasnopolsky R., Li Z.-Y., Blandford R. 1999, ApJ 526, 631\\
Magnetocentrifugal Launching of Jets from Accretion Disks.I- Cold
Axisymmetric Flows 

\reference
Kudoh T., Matsumoto R., Shibata K. 1998, ApJ 508, 186\\
Magnetically Driven Jets from Accretion Disks. III. 2.5-dimensional
Nonsteady Simulations for Thick Disk Case

\reference
Kwan J. 1997, ApJ 489, 284\\
Warm Disk Coronae in Classical T Tauri Stars

\reference
Lery T., Henriksen R.N., Fiege J.D. 1999, A\&A 350, 254\\
Magnetised protostellar bipolar outflows. I. Self-similar model with
Poynting flux 

\reference
Lery T., Heyvaerts J., Appl S., Norman C.A. 1999, A\&A 347, 1055\\
Outflows from magnetic rotators. II. Asymptotic structure and collimation

\reference
Li J. 1996, ApJ 456, 696\\
Magnetic Interaction between Classic T Tauri Stars and Their Associated Disks

\reference 
Li Z.-Y. 1995, ApJ, 444, 848\\
Magnetohydrodynamic disk-wind connection: Self-similar solutions

\reference
Li Z.-Y. 1996, ApJ 465, 855\\ 
Magnetohydrodynamic disk-wind connection: Magnetocentrifugal winds from
ambipolar diffusion-dominated accretion disks

\reference
Livio M. 1997, in IAU 163, Accretion Phenomena and Related Outflows, ASP
Conf Series 121\\ 
The Formation Of Astrophysical Jets

\reference
Lovelace R.V.E. 1976, Nature 262, 649\\
Dynamo model of double radio sources

\reference
Lovelace R.V.E., Wang J.C.L., Sulkanen M.E. 1987, ApJ 62, 1\\
Self-collimated electromagnetic jets from magnetized accretion disks

\reference
Lovelace R.V.E., Romanova M.M., Bisnovatyi-Kogan G.S. 1999, ApJ 514, 368\\
Magnetic Propeller Outflows

\reference
Machida M., Hayashi M.R., Matsumoto R. 2000, ApJ, 532, L67\\
Global Simulations of Differentially Rotating Magnetized Disks: Formation
of Low-$\beta$; Filaments and Structured Coronae

\reference
Mestel L. 1968, MNRAS 138, 359\\
Magnetic braking by a stellar wind-I

\reference
Miller K.A., Stone J.M. 2000, ApJ 584, 398\\
The Formation and Structure of a Strongly Magnetized Corona above a Weakly
Magnetized Accretion Disk

\reference
Mirabel I.F.,Rodriguez L.F. 1999, ARA\&A 37, 409\\
Sources of Relativistic Jets in the Galaxy

\reference
Okamoto I. 1999, MNRAS 307, 253\\
Do magnetized winds self-collimate?

\reference 
Ostriker E. 1997, ApJ 486, 291\\
Self-similar Magnetocentrifugal Disk Winds with Cylindrical Asymptotics

\reference
Ouyed R., Pudritz, R.E. 1999, MNRAS 309, 233\\
Numerical simulations of astrophysical jets from Keplerian discs -III. The
effects of mass loading

\reference
Ouyed R., Pudritz, R.E. 1997, ApJ 484, 794\\
Numerical Simulations of Astrophysical Jets from Keplerian Disks
-II. Episodic Outflows    

\reference
Parker E.N. 1958, ApJ 128, 664\\
Dynamics of the interplanetary gas and magnetic fields

\reference
Pelletier G., Pudritz R.E 1992, ApJ 394, 117\\
Hydromagnetic disk winds in young stellar objects and active galactic nuclei

\reference
Popham R., Narayan R., Hartmann L., Kenyon S. 1993, ApJ 415, L127\\
Boundary Layers in Pre-Main-Sequence Accretion Disks

\reference
Pudritz, R.E., Norman C.A. 1983, ApJ 274, 677\\
Centrifugally driven winds from contracting molecular disks

\reference 
Ray T.P., Mundt R., Dyson J.E., Falle S., Raga A.C. 1996, ApJ 468, L103\\
HST Observations of Jets from Young Stars

\reference
Rekowski M.v., R\"udiger G., Elstner D. 2000, A\&A 353, 813\\
Structure and magnetic configurations of accretion disk-dynamo models

\reference
Rosso F., Pelletier G. 1994, A\&A 287, 325\\
A variational method for solving fast MHD flows. Consequences for stellar
and extragalactic jets

\reference 
Safier P.N. 1993, ApJ 408, 115\\
Centrifugally driven winds from protostellar disks. I - Wind model and
thermal structure 

\reference
Sauty C., Tsinganos K., Trussoni E. 1999, A\&A 348, 327\\
Nonradial and nonpolytropic astrophysical outflows. IV. Magnetic or thermal
collimation of winds into jets?

\reference
Sergeant S., Rawlings S., Maddox S.J., Baker J.C., Clements D., Lacy M.,
Lilje P.B. 1998, MNRAS 294, 494\\
The radio-optical correlation in stepp spectrum quasars

\reference 
Shakura N.I., Sunyaev R.A. 1973, A\&A 24, 337\\
Black holes in binary systems. Observational appearance

\reference 
Shakura N.I., Sunyaev R.A., Zilitinkevich S.S. 1978, A\&A 62, 179\\
On the turbulent energy transport in accretion discs

\reference
Shang H., Shu F.H., Glassgold A.E. 1998, ApJ 493, L91\\
Synthetic Images and Long-slit Spectra of Protostellar Jets

\reference 
Shibata K., Uchida Y. 1985, PASJ, 37, 31\\
A magnetodynamic mechanism for the formation of astrophysical jets. I -
Dynamical effects of the relaxation of nonlinear magnetic twists 

\reference 
Shu F.H., Najita J., Ostriker E., Wilkin F., Ruden S., Lizano S. 1994, ApJ
429, 781\\ 
Magnetocentrifugally driven flows from young stars and disks. I- A
generalized model 

\reference
Shu F.H., Najita J., Ostriker E., Shang H. 1995, ApJ 455, L155\\
Magnetocentrifugally Driven Flows from Young Stars and Disks. V- Asymptotic
Collimation into Jets

\reference
Spruit H.C., Foglizzo T., Stehle R. 1997, MNRAS 288, 333\\
Collimation of magnetically driven jets from accretion discs

\reference 
Stone J.M., Norman M.L. 1994, ApJ 433, 746\\
Numerical simulations of magnetic accretion disks

\reference 
Tagger M., Pellat R. 1999, A\&A 349, 1003\\
An accretion-ejection instability in magnetized disks

\reference
Uchida Y., Shibata K. 1984, PASJ 36, 105\\
Magnetically buffered accretion to a young star and the formation of
bipolar flows 

\reference 
Ustyugova G.V., Koldoba A.V., Romanova M.M., Chechetkin V.M., Lovelace
R.V.E. 1999, ApJ 516, 221\\
Magnetocentrifugally Driven Winds: Comparison of MHD Simulations with Theory

\reference
Ustyugova G.V., Lovelace R.V.E., Romanova M.M., Li H., Colgate S.A. 2000,
ApJ 541, L21\\ 
Poynting Jets from Accretion Disks: Magnetohydrodynamic Simulations

\reference
Vlahakis N., Tsinganos K., Sauty C., Trussoni E. 2000, MNRAS 318, 417\\
A disc-wind model with correct crossing of all magnetohydrodynamic critical
surfaces 

\reference 
Wardle M., K\"onigl A. 1993, ApJ 410, 218\\
The structure of protostellar accretion disks and the origin of bipolar
flows

\end{references}
\end{document}